\documentclass[reprint,amsmath,aps,prb,superscriptaddress]{revtex4-2}

\usepackage{setspace}
\usepackage{lineno}
\usepackage{widetable}
\usepackage{graphicx}
\usepackage{braket}
\usepackage{float}

\usepackage{epstopdf}

\usepackage{dcolumn}
\usepackage{float}
\usepackage{hyperref}
\hypersetup{
	colorlinks   = true, 
	urlcolor     = blue, 
	linkcolor    = blue, 
	citecolor    = blue  
}
\usepackage{placeins}
\usepackage{color}

\usepackage{amsmath}
\usepackage{amssymb}

\usepackage{xspace}
\usepackage{tabularx}
\usepackage{times}

\usepackage{booktabs}

\usepackage{lmodern}

\usepackage{mhchem}

\newcommand{\add}[1]{#1}

\usepackage[english]{babel}

\usepackage{pythonhighlight}

\newcommand{\Ca}{\ce{CaWO4} }

\newcommand{\Er}{\ce{CaWO4}:\ce{Er^3+} }

\newcommand{\etal}{\textit{~et al.}\xspace}

\definecolor{bluegray}{rgb}{0.4, 0.6, 0.8}

\newcommand{\affIMNP}{CNRS, Aix-Marseille Universit\'{e}, Universit\'{e} de Toulon, IM2NP, Marseille, France.}
\newcommand{\affCEA}{Quantronics group, Service de Physique de l'\'Etat Condens\'e  (CNRS, UMR\ 3680),\\IRAMIS, CEA-Saclay, Universit\'e Paris-Saclay, 91191 Gif-sur-Yvette, France}
\newcommand{\affIRCP}{Chimie ParisTech, PSL University, CNRS, Institut de Recherche de Chimie Paris, 75005 Paris, France}

\begin{document}
	
	\title{Electron paramagnetic resonance of \ce{Dy^3+}-doped \ce{CaWO4}: spin Hamiltonian, crystal-field analysis, and linear electric field effect}
	
	\author{Achuthan~Manoj~Kumar}\affiliation{\affIMNP}
	\author{Larissa~Aboudem-Joumessi}\affiliation{\affIMNP}
	\author{Remy~Dassonneville}\affiliation{\affIMNP}
	\author{Adrien~Savoyant}\affiliation{\affIMNP}
	\author{Karolina~Waszowska}\affiliation{\affIRCP}
	\author{Pengrui~Jiao}\affiliation{\affIRCP}
	\author{Patrice~Bertet}\affiliation{\affCEA}
	\author{Philippe~Goldner}\affiliation{\affIRCP}
	\author{Sylvain~Bertaina}\email{sylvain.bertaina@cnrs.fr}\affiliation{\affIMNP}
	
	\begin{abstract}
		We report an electron paramagnetic resonance (EPR) study of \ce{Dy^3+} in
		\ce{CaWO4} single crystals spanning dopant concentrations from 80~ppb to
		247~ppm. The $g$ factors of the ground Kramers doublet at the $S_4$ point-group site,
		\add{$g_\parallel = 7.22 \pm 0.01$ and $g_\perp = 5.45 \pm 0.01$}, and the hyperfine constants
		of \ce{^{161}Dy} and \ce{^{163}Dy} are determined with an order-of-magnitude
		improvement in precision over the only previous report. A crystal-field
		analysis constrained jointly by published optical levels and by the measured
		$g$ tensor identifies the ground doublet as the lower branch of an
		anticrossing between the near-degenerate $\ket{{\pm}11/2}$ and
		$\ket{{\mp}13/2}$ states, and explains why optical data alone left the $g$
		values unconstrained. \add{The same parameter set, tested on the excited
			multiplets of the $^6H$ term, reproduces the ${}^6H_{13/2}$ crystal-field
			levels better than the optical parameters themselves.} The angular dependence of the linewidth in the $ab$
		plane reveals broadening by random internal electric fields through the
		linear electric field effect. Calibrating these fields with the residual
		\ce{Er^3+} present in the same crystals yields the first electric-field
		coupling parameters of \ce{Dy^3+},
		$(B_{31}^2+B_{36}^2)^{1/2} = (34\pm5)\times10^{-6}$~cm/V, three times the
		\ce{Er^3+} value and the largest reported for a rare-earth ion in this host.
		The residual linewidth and the concentration-dependent lineshape asymmetry
		follow the quadratic field scaling expected for a second-order Stark shift,
		which escapes the inversion-site cancellation constraining the linear
		effect. Both couplings trace back to the $\sim 20$~cm$^{-1}$ gap to the first
		opposite-symmetry doublet, a consequence of the dense crystal-field
		structure of the $J=15/2$ manifold.
	\end{abstract}
	
	\pacs{76.30-v, 71.70.Ch, 71.70.Ej}

	\maketitle
	
	\section{Introduction}
	
	\Ca holds a special place in the history of electron paramagnetic resonance (EPR) spectroscopy. The
	pioneering work of Mims \etal during the 1960s established this scheelite-type crystal as a
	foundational system for the development of pulsed EPR techniques that continue to shape the field
	today. Effects like spectral diffusion\cite{mims1961}, electron spin echo envelope modulation
	(ESEEM) \cite{rowan1965} and electrical field shift \cite{mims1964,mims1965,mims1966} have been
	observed and modeled using \Ca.
	
	Interest in rare-earth doped \Ca was renewed by quantum technology applications, with
	the observation of long, coherent Rabi oscillations in a rare-earth solid-state system using \Er at liquid helium
	temperature \cite{bertaina2007,bertaina2009}. More recently, the \Ca platform has enabled
	demonstrations in hybrid quantum systems. Thanks to the Purcell effect, EPR spectroscopy in the
	millikelvin range was demonstrated \cite{bienfait2016}, enabling a Hahn-echo coherence time $T_2$
	of 23~ms for \ce{Er^3+} electron spins \cite{ledantec2021,rancic2022}. Ultimately, single electron
	spins \cite{wang2023,billaud2025} and then single $^{183}$W nuclear spins \cite{osullivan2025,travesedo2025}
	were detected in \Er using microwave fluorescence detection. The platform
	has since been extended beyond erbium, with sub-second optical and spin coherences recently
	reported for \ce{Yb^3+}:\Ca \cite{tiranov2026}.
	
	Despite this extensive work spanning six decades \cite{hempstead1960,forrester1962,kirton1964,kirton1965,kiel1970}, EPR of \ce{Dy^3+} in \Ca remains largely
	unstudied. The 1966 paper by Antipin \etal \cite{antipin1966} represents a pioneering but
	preliminary observation of \ce{Dy^3+} EPR in \Ca. While historically significant as the first report
	of such measurements, the work suffers from limitations: large parameter uncertainties, absence of
	angular dependence data, lack of spectral figures, and incomplete analysis of line broadening
	mechanisms. These deficiencies made a more detailed investigation not only desirable but essential
	for establishing reliable spin Hamiltonian parameters and understanding the electronic structure of
	\ce{Dy^3+} in the scheelite lattice.
	
	In this work, we determine the spin-Hamiltonian parameters of the $S_4$ \ce{Dy^3+} center with
	substantially improved precision, reconcile them with the published optical crystal-field data
	through a joint fit, and use the angular dependence of the linewidth to extract the first linear
	electric-field coupling parameters of \ce{Dy^3+}, together with evidence for a second-order Stark
	contribution to the lineshape.

	\section{Experimental Details}
	
	\subsection{Crystal structure}
	
	\Ca crystallizes in the tetragonal scheelite structure, space group $I4_1/a$ (no.~88), with
	four formula units per unit cell and lattice constants $a=b=5.24$~\AA, $c=11.37$~\AA\
	\cite{ledantec2021,hazen1985}. The structure contains two inequivalent cation sites: the \ce{W^6+} site,
	fourfold coordinated by oxygen within a rigid \ce{WO4^2-} tetrahedron, and the \ce{Ca^2+} site,
	eightfold coordinated, whose point symmetry is $S_4$. Comparison of ionic radius and charge
	places \ce{Dy^3+} on the \ce{Ca^2+} site rather than the \ce{W^6+} site
	\cite{wortman1971,antipin1966}, so the dopant inherits the $S_4$ point symmetry of the host site.
	This is the symmetry that constrains the form of the spin Hamiltonian and of the local electric
	field tensor used throughout Sec.~\ref{sec:LEFE}.
	
	Substitution of \ce{RE^3+} for \ce{Ca^2+} leaves a charge excess of $+1$ per dopant ion. This
	excess is compensated at long range elsewhere in the lattice, most commonly by 
	\ce{Na+} introduced during growth \cite{wortman1971}. The compensating charge is randomly  distributed in the crystal,  if close to  
	\ce{RE^3+}, the local symmetry change is lowered giving orthorhombic site but mostly the compensation charge is far from \ce{RE^3+}
	and the symmetry remains $S_4$ on average; its random position from one dopant to the next is, however, precisely what generates the
	distribution of local electric fields discussed in Sec.~\ref{sec:LEFE}.
	
	\subsection{Crystal growth and sample preparation}
	
	Three single crystals, labeled (s1), (s2), and (s3), were used in this study, with a typical
	size of $2\times2\times3$~mm$^3$. All three originate from boules grown by the Czochralski
	method from high-purity \ce{CaCO3} and \ce{WO3} powders \cite{wang2023,ledantec2021,nassau1963,brissot1964}.
	
	Sample (s1) was pulled from an undoped melt: no \ce{Dy^3+} or \ce{Na+} was intentionally added.
	The trace \ce{Dy^3+} content detected in this sample originates from naturally occurring
	impurities in the starting materials, at the level reported in Table~\ref{tab:samples}.
	
	Samples (s2) and (s3) originate from two independent growth batches, available in the laboratory
	stock for several years prior to this study. In both cases \ce{Dy^3+} was added directly to the
	melt as the intentional dopant, together with \ce{Na+} co-doping for charge compensation. Because
	these crystals predate the present study, the exact melt composition and growth protocol used for
	each batch are not documented;  the nominal doping levels quoted in Table~\ref{tab:samples} are
	the values reported at the time of growth, and the actual incorporated concentration was
	therefore determined independently by EPR for this work.
	
	The crystallographic axes $a$, $b$, $c$ were identified by X-ray (Laue) diffraction, with the
	sample subsequently aligned to better than $2^{\circ}$.
	
	The actual \ce{Dy^3+} concentration in each sample was determined by continuous-wave EPR, using
	double integration of the absorption-derivative signal and comparison against a \Er reference
	crystal of independently calibrated concentration \cite{ledantec2021}. The measured
	concentrations, hereafter denoted $[\mathrm{Dy}^{3+}]$, are reported in Table~\ref{tab:samples}
	together with the nominal melt doping.
	
	\begin{table}[h]
		\caption{Nominal and EPR-calibrated \ce{Dy^3+} concentrations for the three samples studied.}
		\label{tab:samples}
		\begin{tabular}{cccc}
			\toprule
			Sample & Na co-doping & Nominal doping & $[\mathrm{Dy}^{3+}]$ (measured) \\
			\midrule
			(s1) & no  & --      & 80~ppb \\
			(s2) & yes & 10~ppm   & 2.3~ppm \\
			(s3) & yes & 2000~ppm & 247~ppm \\
			\bottomrule
		\end{tabular}
	\end{table}
	
	\subsection{EPR Setup}
	
	EPR measurements were performed on a conventional Bruker EMX X-band spectrometer. A standard
	rectangular cavity equipped with optical access was used, excited in the TE$_{102}$ mode at a
	resonance frequency of 9.39\,GHz. Samples were mounted on a Suprasil quartz holder and cooled in an
	ESR~900 cryostat coupled to a Cold Edge Stinger cryogen-free cryocooler, reaching a base temperature
	of 7~K; no saturation of the EPR signal was observed over the accessible temperature range even at maximum microwave power.
	The modulation frequency was \add{100~kHz} and the amplitude up to 10~G (below the minimum linewidth). The
	orientation of the crystal with respect to the static magnetic field, $\theta$,  was set by an automated
	goniometer, with an angular error below $0.5^\circ$. Between successive measurement series, the
	goniometer was reset to $0^\circ$ to guarantee reproducibility of the angular dependence.
	
	\section{Results and Discussion}

	\subsection{Spin Hamiltonian}
	\label{sec:spin-hamiltonian}
	
	The ground state of \ce{Dy^3+} in \Ca\ is the lowest Kramers doublet
	\add{of the $^{6}H_{15/2}$ manifold, split off from the rest} by the tetragonal
	($S_4$) crystal field. \add{It is isolated in the sense that matters for
		X-band EPR: no other doublet contributes a resolved resonance, and the
		spectra of Fig.~\ref{fig:PG2} are accounted for by a single effective spin.
		The crystal-field analysis of Appendix~\ref{app:CF} nevertheless places a
		second doublet only a few cm$^{-1}$ higher, appreciably populated at 7~K;
		no separate resonance is resolved from it, as expected if its lifetime is
		limited by the same small gap.} Within this
	doublet the system behaves as an
	effective spin $S=1/2$, and, following the notation of
	Antipin~\etal~\cite{antipin1966}, the EPR spectrum of the odd isotopes
	\ce{^{161}Dy} and \ce{^{163}Dy} ($I=5/2$) is described by the axial spin
	Hamiltonian
	
	\begin{equation}
		\mathcal{H} = g_\parallel \mu_B H_z S_z + g_\perp \mu_B (H_x S_x + H_y S_y)
		+ A_\parallel I_z S_z + A_\perp (I_x S_x + I_y S_y),
		\label{eq:spinham}
	\end{equation}
	
	where $z$ is taken along the crystallographic $c$ axis, $\mu_B$ is the Bohr
	magneton, \add{$\mathbf{H}$ denotes the magnetic induction at the sample
		,}
	and $g_\parallel$, $g_\perp$ are the principal $g$-values imposed
	by the $S_4$ site symmetry. $A_\parallel$ and $A_\perp$ are the axial
	components of the hyperfine tensor. The nuclear quadrupole interaction,
	although allowed for $I=5/2$, is omitted from Eq.~(\ref{eq:spinham}): at
	X band it shifts the allowed $\Delta m_I = 0$ transitions only at second
	order and is not resolved within the linewidths reported below. For the
	even isotopes \ce{^{160}Dy}, \ce{^{162}Dy} and \ce{^{164}Dy} ($I=0$,
	combined natural abundance 56.1\%), the last two terms of
	Eq.~(\ref{eq:spinham}) vanish identically, and the spectrum reduces to the
	single line set by the electronic Zeeman term alone, which accounts for the
	central resonance in Figure~\ref{fig:PG2}.
	
	\subsection{EPR spectroscopic parameters}
	\label{sec:epr-params}

	\begin{figure}[h]
		\includegraphics[width=\columnwidth]{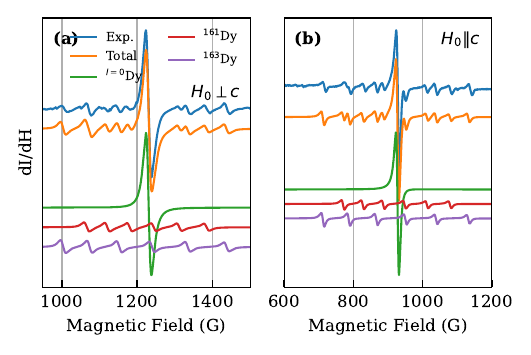}
		\caption{\ce{Dy^3+} EPR spectra in \Ca\ sample (s1), recorded at 7~K with
			$H_0 \perp c$ (a) and $H_0 \parallel c$ (b) (blue curves). The orange
			curve is the total simulated spectrum, built from the
			abundance-weighted sum of the $I=0$ (green), \ce{^{161}Dy} (red) and
			\ce{^{163}Dy} (purple) sub-spectra, each calculated with the
			spin-Hamiltonian parameters given in Eq.~\eqref{eq:fitvalues}, \add{each
				transition being described by a Dysonian lineshape}.}
		\label{fig:PG2}
	\end{figure}

	Figure~\ref{fig:PG2} shows representative \ce{Dy^3+} EPR spectra recorded
	at 7~K with $H_0 \perp c$ (a) and $H_0 \parallel c$ (b). The spectrum is dominated by a single intense
	resonance flanked by a series of weaker satellites. This pattern is the fingerprint of the natural isotopic
	composition of dysprosium. Three even isotopes, \ce{^{160}Dy},
	\ce{^{162}Dy} and \ce{^{164}Dy}, carry no nuclear spin ($I=0$) and together
	account for 56.1\% of natural Dy; they contribute the central line, free of
	hyperfine structure. The two odd isotopes, \ce{^{161}Dy} (18.9\%) and
	\ce{^{163}Dy} (24.9\%), both possess $I=5/2$ and split into a sextet of
	hyperfine lines through their respective hyperfine tensors. The simulated
	spectrum, obtained as the abundance-weighted sum of the three isotopologue
	contributions calculated with \texttt{EasySpin}~\cite{stoll2006}, reproduces
	both the central feature and the satellite positions, allowing an
	unambiguous assignment of the latter to the \ce{^{161}Dy} and \ce{^{163}Dy}
	hyperfine manifolds.
	
	We note that the \ce{Dy^3+} EPR line, central feature and satellites alike,
	is asymmetric. In order to extract the resonance fields with the required
	accuracy, we described each line by a Dysonian function rather than by a
	pure Lorentzian (or Gaussian) derivative, so as to account for the shift in
	resonance field induced by this lineshape asymmetry. A potential physical
	interpretation of this asymmetry is given later in the text; we can
	already state that it cannot be attributed to a skin-depth effect\add{, the
		origin of the Dysonian profile in conduction-electron resonance}, as
	observed in conducting samples, since \Ca\ is a very good insulator.
	
	The best-fit spin-Hamiltonian parameters, obtained by simultaneously
	adjusting the simulated spectra of \ce{^{161}Dy} and \ce{^{163}Dy} to the
	data of Figure~\ref{fig:PG2}, are

	\begin{equation}
		\left.
		\begin{aligned}
			g_\parallel &= \add{7.22 \pm 0.01}, & g_\perp &= \add{5.45 \pm 0.01},\\
			A_\parallel^{161} &= 625 \pm 5~\mathrm{MHz}, &
			A_\perp^{161} &= 485 \pm 5~\mathrm{MHz},\\
			A_\parallel^{163} &= 850 \pm 5~\mathrm{MHz}, &
			A_\perp^{163} &= 660 \pm 5~\mathrm{MHz},\\
			\frac{A_\parallel^{161} g_\perp}{A_\perp^{161} g_\parallel} &= 0.973 \pm 0.020, &
			\frac{A_\parallel^{163} g_\perp}{A_\perp^{163} g_\parallel} &= 0.973 \pm 0.020.
		\end{aligned}
		\right\}
		\label{eq:fitvalues}
	\end{equation}
	
	\add{The uncertainties in Eq.~(\ref{eq:fitvalues}) are not set by the
		linewidth itself. Each spectrum contains thirteen transitions, the $I=0$
		central line together with the two $I=5/2$ sextets, whose positions are
		returned by a single global fit; the centre of a resonance is located far
		more accurately than its width once the lineshape is modelled. In field
		units, $\pm0.01$ on $g_\parallel$ amounts to $\pm1.3$~G at a resonance
		field of $929$~G, about one fifteenth of the 20~G width, and $\pm0.01$ on
		$g_\perp$ to $\pm2.3$~G at $1231$~G. Crystal misalignment contributes
		little: at a principal axis the effective $g$ varies quadratically with the
		tilt angle, so a $2^\circ$ error displaces $g_\parallel$ by $0.002$ and
		$g_\perp$ by $0.003$, and the error is further suppressed by fitting the
		complete $ac$-plane rotation of Fig.~\ref{fig:WFPG2}(a) rather than two
		isolated settings. The limiting term is the lineshape model, since the
		asymmetry discussed above displaces the apparent centre of a symmetric
		derivative by several gauss. We therefore quote $\pm0.01$, which brackets
		the values returned when the Dysonian phase is varied across the range
		compatible with the data, and refrain from a third decimal that this budget
		would not support.}
	
	The ratio $A_\parallel g_\perp / A_\perp g_\parallel$ deserves comment.
	Within a Kramers doublet belonging to a single $J$ multiplet, the Zeeman
	and hyperfine interactions are both projections of vector operators, and
	the Wigner--Eckart theorem forces $A_\parallel/g_\parallel =
	A_\perp/g_\perp$, a consequence of the projection theorem: within an
	isolated $J$ multiplet, both the hyperfine and Zeeman interactions reduce
	to scalar multiples of the same operator $\mathbf{J}$, so their tensors
	are forced proportional \cite{abragam2013}, i.e., a ratio of unity, whatever the crystal-field
	composition of the doublet; small deviations may also arise from
	non-dipolar hyperfine contributions such as core polarization\add{, the
		polarization of the filled inner $s$ shells by the $4f$ moment} (the
	Fermi contact, or point-contact, term), expected to
	be weak for $4f$ ions. The measured value, \add{$0.973 \pm 0.020$ for both
		isotopes, is quoted with an uncertainty deliberately larger than the
		quadrature propagation of Eq.~(\ref{eq:fitvalues}), which returns $0.013$
		for \ce{^{161}Dy} and $0.010$ for \ce{^{163}Dy}; the excess covers the
		lineshape systematics common to all four hyperfine constants, which do not
		average down. It is}
	a direct experimental bound on the $J$-mixing of the ground
	doublet, consistent with the $>99.9\%$ $^6H_{15/2}$ purity found in the
	crystal-field analysis of Appendix~\ref{app:CF}. \add{The same ratio serves
		as a $J$-mixing diagnostic in Kirton's compilation for this host, where it
		is $0.97$ for \ce{Nd^3+} and $0.38$ for \ce{Sm^3+}, the latter taken as
		evidence of substantial admixture \cite{kirton1965}.} As a further internal check, the ratio $A^{163}/A^{161}=1.360$, identical for
	the parallel and perpendicular components, reproduces the ratio of the nuclear
	$g$ factors of the two isotopes ($\simeq1.40$) to within 3\%. We note that the
	$\pm5$~MHz uncertainties of Eq.~(\ref{eq:fitvalues}), which reflect the
	statistical scatter of the fit alone, would make this a three-standard-deviation
	departure; the residual is more plausibly ascribed to systematic error in the
	lineshape model than to a hyperfine anomaly, which is \add{of order
		$10^{-3}$ for this isotope pair \cite{abragam2013}}.

	Two literature determinations of the electronic $g$-tensor are available
	for comparison. Antipin~\etal~\cite{antipin1966} measured
	$g_\parallel = 7.5 \pm 1.0$ and $g_\perp = 5.5 \pm 0.5$ from EPR at
	$\sim$3~GHz on a crystal doped to $\sim$1\% Dy, \add{three to four orders of
		magnitude above the trace concentration used here}. Given this operating
	frequency and doping level, the determination is a creditable one, and it
	agrees with our result throughout. \add{A second pair of values,
		$g_\parallel = 7.267$ and $g_\perp = 5.466$, appears in the review table of
		Kirton~\cite{kirton1965}, where it is attributed to his own unpublished
		work rather than to a separate paper. The comparison is instructive. Our
		$g_\perp$ reproduces his to within $0.016$, well inside the combined
		uncertainty, whereas his $g_\parallel$ exceeds ours by $0.047$, five times
		the error bar of Eq.~(\ref{eq:fitvalues}). A discrepancy confined to one
		principal value cannot be blamed on field calibration, which would rescale
		both components in the same proportion; it points to the determination of
		$g_\parallel$ specifically, that is, to the orientation for which
		Fig.~\ref{fig:conc}(b) shows the lineshape asymmetry to be strongest and a
		symmetric derivative model to be least adequate.}
	
	\add{Whether the two results are in fact in conflict cannot be established.
		The tabulated entry carries no uncertainty, no spectrum, no linewidth and
		no hyperfine constants, and the row itself records that the available data
		were insufficient to determine the ground-state eigenvector
		\cite{kirton1965}. What can be said is that the concentration series
		reported here bears on the question. The residual width $\Gamma_0$ measured
		at the linewidth-minimizing azimuth rises from $20$~G at 80~ppb to $202$~G
		at 247~ppm (Sec.~\ref{sec:LEFE2}), and the \Ca\ crystals used throughout
		Ref.~\cite{kirton1965} were pulled from melts containing $0.1\%$ rare
		earth, the figure quoted there. Lines an order of magnitude broader than
		ours, recorded without a lineshape correction for the asymmetry, are
		unlikely to sustain a determination at the third decimal place. We take the
		agreement on $g_\perp$ as the meaningful comparison and regard the
		$g_\parallel$ difference as a measure of the systematic error attached to
		an undocumented measurement, rather than as evidence of a real difference
		between the two crystals.}
	
	On the hyperfine side, no constants accompany Kirton's unpublished
	$g$-values. Antipin~\etal, by contrast, extracted $A_\parallel$ and
	$A_\perp$ for both \ce{^{161}Dy} and \ce{^{163}Dy}: converted from
	their tabulated values in units of $10^{-3}~\mathrm{cm}^{-1}$,
	$A_\parallel^{161} = 660\pm90$~MHz, $A_\perp^{161} = 480\pm60$~MHz,
	$A_\parallel^{163} = 929\pm120$~MHz, and $A_\perp^{163} = 690\pm90$~MHz; despite the
	sizeable uncertainties attached to these four values, of order
	60--120~MHz, they constitute a sound determination for the frequency and
	concentration regime in which they were obtained, and they agree with our
	result within their respective error bars (cf.\ Eq.~\eqref{eq:fitvalues}). By presenting the raw spectra
	alongside their simulation, the present work provides $g$-factors and
	hyperfine couplings with a substantially improved relative precision,
	\add{$\pm0.01$ on $g$ and $\pm5$~MHz on the hyperfine constants, together
		with an explicit account of the systematics that limit them}.
	
	\subsection{Reconciling optics and EPR}
	\label{sec:reconciling}
	Crystal-field calculations of the ground-doublet $g$ tensor are available
	for comparison. Diagonalizing the crystal field with the $B_{kq}$ that
	Wortman and Sanders fitted to their optical spectra \cite{wortman1971},
	we obtain $g_\parallel = 12.4$, $g_\perp = 2.4$, in agreement with the
	values quoted in Ref.~\cite{wortman1971} itself\add{, whose $g_\parallel$
		carries a negative sign that a cw EPR experiment cannot access, since the
		resonance condition involves $|g|$ alone}, but far from the EPR
	measurement. A later
	diagonalization from the same $B_{kq}$ \cite{yang2010} reports
	$g_\parallel = 7.62$, $g_\perp = 4.00$; we could not reproduce these
	numbers from the inputs stated there, and in any case $g_\perp$ still
	lies far from our measurement. Wortman and Sanders had in fact already
	located the difficulty. Scanning parameter sets of equal optical
	quality, they found $g_\parallel$ anywhere between $-12$ and $+12$
	\cite{wortman1971}: the optical energies leave the composition of the
	ground doublet, and with it the entire $g$ tensor, essentially
	unconstrained. Any $g$ value computed from an optical-only parameter
	set inherits this indeterminacy.
	
\begin{figure}[tb]
		\includegraphics[width=\columnwidth]{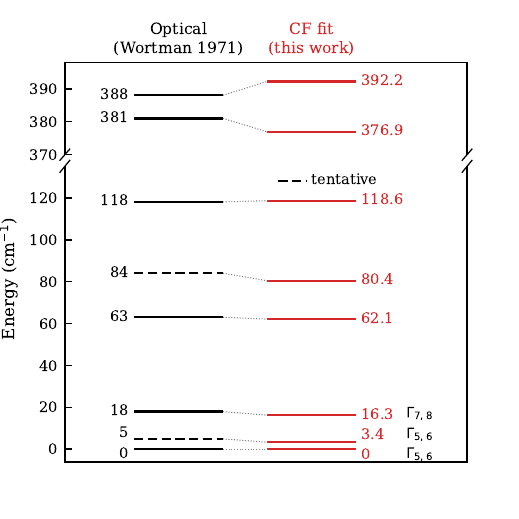}
		\caption{\label{fig:levels}Crystal-field level scheme of the $^6H_{15/2}$
			ground multiplet of Dy$^{3+}$ at the $S_4$ site of CaWO$_4$. Left: the eight
			Kramers doublets located optically by Wortman and Sanders~\cite{wortman1971};
			dashed lines mark their two tentative assignments, at 5 and 84~cm$^{-1}$.
			Right: the levels returned by the joint fit of Eq.~(\ref{eq:Bkq-final}), carried out
			on the complete $^6H$ term and constrained simultaneously by these optical
			energies and by the measured $g$ tensor (\textcolor{black}{rms 2.6}~cm$^{-1}$,
			\textcolor{black}{the two largest residuals falling on the upper doublets near
				385~cm$^{-1}$}). The irreducible
			representation of the three lowest doublets is indicated. The first excited
			doublet, at \textcolor{black}{3.4}~cm$^{-1}$, belongs to the same $\Gamma_{5,6}$ representation
			as the ground state and is connected to it only through the $\Delta M=\pm4$
			chain; the doublet at \textcolor{black}{16.3}~cm$^{-1}$ is the lowest state of the opposite
			representation $\Gamma_{7,8}$. As shown in Sec.~\ref{sec:LEFE-origin}, the two
			gaps govern two \emph{different} electric-field coupling channels: the
			$\Gamma_{7,8}$ gap controls the response to $\mathbf{E}\parallel c$, the
			$\Gamma_{5,6}$ gap that to $\mathbf{E}\perp c$.}
	\end{figure}

	The answer is to constrain the crystal field by both data sets at once.
	We proceed in two steps detailed in
	Appendix~\ref{app:CF}. In a first step, the crystal field is
	diagonalized within the ${}^6H_{15/2}$ multiplet alone (16 states), and
	the five $S_4$ parameters $(B_{20}, B_{40}, B_{60}, B_{44},
	\mathrm{Re}\,B_{64})$ are adjusted simultaneously to the eight optical
	levels of Ref.~\cite{wortman1971} and to the measured $(g_\parallel,
	g_\perp)$. \textcolor{black}{Started from one hundred random points of the
		five-dimensional parameter space, the search converges to a single best
		solution, which returns both $g$ values to better than $0.01$ and the
		eight optical energies to $2.5~\mathrm{cm}^{-1}$ rms. This settles the
		main point: a parameter set compatible with the optical spectrum and
		with the EPR data does exist, and the two data sets are not in
		conflict.} In
	a second step, the diagonalization is extended to the complete $^6H$
	term ($J = 15/2, \ldots, 5/2$, 66 states), so that the $J$-mixing
	induced by the crystal field is treated exactly; the 16-state solution
	serves as the starting point of this final fit. The result,
	\begin{equation}
		\begin{aligned}
			B_{20} &= \textcolor{black}{449}, & B_{40} &= \textcolor{black}{-919}, & B_{60} &= \textcolor{black}{20},\\
			B_{44} &= \textcolor{black}{812}, & \mathrm{Re}\,B_{64} &= \textcolor{black}{662} \quad (\mathrm{cm}^{-1}),
		\end{aligned}
		\label{eq:Bkq-final}
	\end{equation}
	returns both $g$ values and reproduces the optical levels to
	\textcolor{black}{$2.6~\mathrm{cm}^{-1}$} rms, below the $9~\mathrm{cm}^{-1}$ scatter of
	the optical-only fit \cite{wortman1971}: the $g$ constraint costs
	nothing in optical residual. It selects one point along a direction the
	optical energies leave flat rather than competing with them.
	\textcolor{black}{The two steps agree on every observable and differ by a
		few percent on $B_{20}$, $B_{40}$ and $B_{44}$; the exception is
		$B_{60}$, which changes sign between them and is the least determined of
		the five.}
	\add{The resulting level scheme is shown in Fig.~\ref{fig:levels}. The
		parameter set can be tested further on data that entered neither
		fit. The seven crystal-field levels of the ${}^6H_{13/2}$ multiplet,
		located optically in the same study, are reproduced to
		\textcolor{black}{$6.1~\mathrm{cm}^{-1}$} rms, against $9.8~\mathrm{cm}^{-1}$ for the
		optical-only parameters evaluated in the same model
		(Appendix~\ref{app:CF-multiplets}). The $g$-constrained set is thus the
		better description of a multiplet it never saw.}
	
	\add{One piece of notation is needed before proceeding. Under $S_4$ the
		projection $M$ is conserved only modulo 4, so the eight Kramers doublets of
		${}^6H_{15/2}$ fall into two families that the crystal field never
		connects: those built from $M = \pm 11/2, \pm 3/2, \mp 5/2, \mp 13/2$,
		labelled $\Gamma_{5,6}$, and those built from the complementary set,
		labelled $\Gamma_{7,8}$. The ground doublet belongs to the first. This
		partition is immaterial for the level positions, but it turns out to
		control which excited doublet an electric field can reach
		(Sec.~\ref{sec:LEFE-origin}).}
	
	The resulting ground-state wavefunction, $\psi_+ \simeq \textcolor{black}{0.805}
	\,|\textcolor{black}{13/2}\rangle - \textcolor{black}{0.536}\,|{\textcolor{black}{-11/2}}\rangle$ plus small \textcolor{black}{$|5/2\rangle$} and
	\textcolor{black}{$|{-3/2}\rangle$} admixtures, carries a weight outside ${}^6H_{15/2}$
	of $4\times10^{-4}$. This is the quantity the hyperfine ratio of
	Eq.~(\ref{eq:fitvalues}) bears on. Within a single multiplet the
	projection theorem makes $g$ and $A$ proportional to the same
	matrix elements of $\mathbf{J}$, so that $A_\parallel g_\perp /
	A_\perp g_\parallel = 1$ whatever the crystal-field composition of
	the doublet; the ratio is blind to the \textcolor{black}{$|13/2\rangle$} and
	\textcolor{black}{$|{-11/2}\rangle$} amplitudes and probes only the admixture of other
	$J$ values, for which the reduced matrix elements of the Zeeman and
	hyperfine operators no longer stand in the same proportion. Since
	the departure from unity is first order in the admixture
	\emph{amplitude}, a calculated weight of $4\times10^{-4}$
	corresponds to amplitudes near $2\times10^{-2}$ and to an expected
	deviation of a few percent, which the measured $0.973 \pm 0.020$
	accommodates.

	The fit places a second doublet of the same
	composition inverted, \textcolor{black}{$0.83\,|11/2\rangle + 0.55\,|{-13/2}\rangle$}, at
	\textcolor{black}{$3.4~\mathrm{cm}^{-1}$} above the ground state ($2$--$5~\mathrm{cm}^{-1}$
	along the fit valley), with predicted \textcolor{black}{$g_z \approx 5.1$ and $g_\perp
		\approx 6.4$}, a prediction open to test by far-infrared or
	high-frequency EPR spectroscopy.
	
	\subsection{Angular dependence}
	\label{sec:angular}
	
	\begin{figure}[!htb]
		\includegraphics[width=\columnwidth]{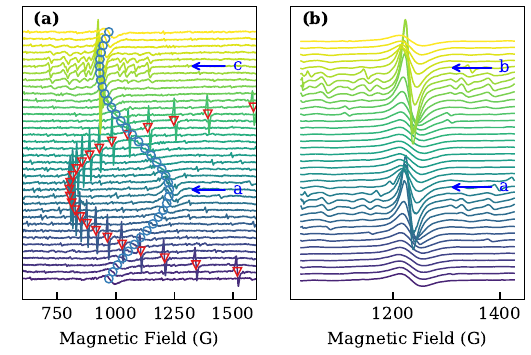}
		\caption{Angular dependence of the Dy$^{3+}$:CaWO$_4$ EPR spectra in
			sample (s1) (80~ppb). (a) Rotation in the $ac$ plane. Blue circles: calculated
			Dy$^{3+}$ resonance fields using the effective $g$-factors determined
			previously; red squares: calculated resonance fields for residual
			Er$^{3+}$ impurities ($\sim$1.5~ppb). (b) Rotation in the $ab$ plane. The
			$S_4$ symmetry imposes an isotropic effective $g$-factor, but the
			linewidth varies strongly with angle, reaching a minimum at
			$\varphi \approx 24^{\circ}$ from the $a$ axis rather than along $a$ or
			$b$.}
		\label{fig:WFPG2}
	\end{figure}

	Figure~\ref{fig:WFPG2} presents the angular dependence of the EPR spectra
	recorded for sample (s1) (80~ppb). Panel (a) shows the rotation pattern in the $ac$
	plane: the blue circles indicate the calculated resonance fields of
	Dy$^{3+}$ in CaWO$_4$, obtained using the effective $g$-factors determined
	previously, while the red squares mark the expected resonance positions
	of residual Er$^{3+}$ impurities, whose concentration in (s1) is estimated
	at $\sim$1.5~ppb.
	
	Panel (b) shows the rotation pattern in the $ab$ plane. As required by
	the $S_4$ site symmetry, the resonance field, i.e., the effective
	$g$-factor, is isotropic in this plane. However, the linewidth
	exhibits a pronounced angular dependence. Notably, the narrowest linewidth does not occur when the field is along the
	crystallographic $a$ or $b$ axis, but at an angle $\varphi \approx 24^{\circ}$
	from $a$, a signature of the linear electric-field effect (LEFE) at the $S_4$
	site, characterized quantitatively in Sec.~\ref{sec:LEFE}. The $ab$-plane
	rotation also reveals additional resonances arising from non-$S_4$
	Dy$^{3+}$ sites, which we assign to orthorhombic sites associated with charge
	compensation.

	\begin{figure}[!htb]
		\includegraphics[width=\columnwidth]{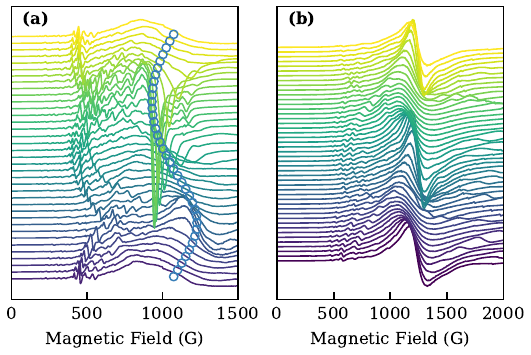}
		\caption{Angular dependence of the Dy$^{3+}$:CaWO$_4$ EPR spectra in the
			high-concentration sample (s3) (247~ppm) recorded at 7~K. (a) Rotation in the $ac$ plane.
			(b) Rotation in the $ab$ plane. In addition to the main $S_4$ Dy$^{3+}$
			line, intense low-field satellite resonances are observed, growing with
			Dy$^{3+}$ concentration and showing pronounced anisotropy in the $ab$
			plane incompatible with $S_4$ symmetry. These signals, first reported
			without analysis by Antipin~\textit{et al.}~\cite{antipin1966}, are
			attributed to Dy$^{3+}$-Dy$^{3+}$ pair centers.}
		\label{fig:WF2000ppm}
	\end{figure}
	
	Figure~\ref{fig:WF2000ppm} shows the angular dependence of the EPR
	spectra for the highest-concentration sample (s3) (247~ppm
	Dy$^{3+}$ doping), with the $ac$-plane rotation in panel (a) and the
	$ab$-plane rotation in panel (b). As in sample (s1), the main $S_4$ Dy$^{3+}$
	line is clearly identified and follows the expected angular dependence
	of the effective $g$-factor. However, at this higher concentration the
	non-$S_4$ satellite lines are considerably more intense and more
	numerous than in (s1), and we assign them to Dy$^{3+}$-Dy$^{3+}$
	pair centers, whose statistical weight scales with the square of the
	dopant concentration.
	
	These low-field satellite signals, appearing well below the main
	resonance, were already noted by Antipin~\textit{et al.}~\cite{antipin1966}
	in their early EPR study of Dy$^{3+}$:CaWO$_4$, but no analysis of their
	microscopic origin was given at that time. Our angular-dependent
	measurements in the $ab$ plane show that these additional lines are
	distinctly anisotropic and incompatible with $S_4$ site symmetry,
	ruling out a simple crystal-field or charge-compensation origin
	analogous to the orthorhombic sites discussed above for (s1). The marked
	growth of their intensity with Dy$^{3+}$ concentration, together with
	this symmetry-breaking angular behavior, supports their identification
	as exchange- or dipolar-coupled Dy$^{3+}$ pairs, whose resonance
	condition is shifted to lower field by the intra-pair interaction.
	
	A simple estimate of the relative intensity of the pair satellite lines
	supports this assignment. In CaWO$_4$, each Dy$^{3+}$ ion has four
	nearest Ca$^{2+}$ neighbors available for substitution. The probability
	that at least one of these four sites is also occupied by a Dy$^{3+}$
	ion is $P_{\mathrm{pair}} = 1-(1-c)^4$, where $c$ is the Dy$^{3+}$
	doping fraction. The four nearest-neighbor directions define four pair
	orientations, magnetically inequivalent for a general orientation of
	$H_0$ in the $ab$ plane, so a single pair line carries one quarter of
	the total pair intensity. Normalizing to the intensity of an isolated
	$S_4$ ion gives $I_{\mathrm{pair}}/I_{\mathrm{single}}
	= P_{\mathrm{pair}}/4 \approx c$ for $c \ll 1$. For sample (s3)
	($c = 247$~ppm), this yields $I_{\mathrm{pair}}/I_{\mathrm{single}}
	\approx 2.5\times10^{-4}$, in quantitative agreement with the experimental
	ratio of a few $10^{-4}$ estimated from the satellite-to-main-line
	intensity ratio (see Appendix~\ref{app:pairs} for details). For the
	lower-concentration samples (s1) and (s2), the same model predicts
	$I_{\mathrm{pair}}/I_{\mathrm{single}} \approx 8\times10^{-8}$ and
	$2.3\times10^{-6}$, respectively, well below the experimental
	detection threshold, consistent with the absence of resolved pair
	satellites in these spectra.

	\begin{figure}[h]
		\includegraphics[width=\columnwidth]{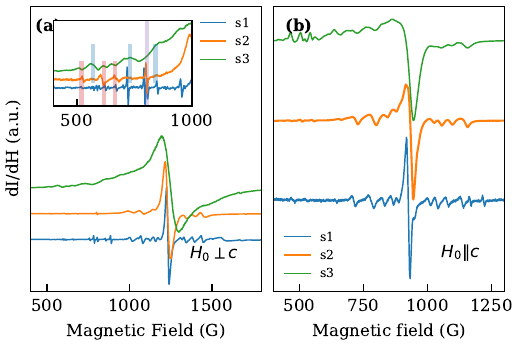}
		\caption{\ce{Dy^3+} EPR spectra in \Ca\ at 7~K for increasing dopant
			concentration. (a) $H_0 \perp c$, recorded at the linewidth-minimizing
			azimuth $\varphi \approx 24^{\circ}$ from $a$ (Fig.~\ref{fig:WFPG2});
			(b) $H_0 \parallel c$. Blue, orange, and green traces correspond to
			samples (s1, 80~ppb), (s2, 2.3~ppm), and (s3, 247~ppm). The linewidth
			and lineshape asymmetry both increase with concentration, the
			asymmetry being far more pronounced for $H_0 \parallel c$, where the
			growing linewidth also blends the \ce{^{161}Dy} and \ce{^{163}Dy}
			hyperfine satellites into the central line in (s3); for
			$H_0 \perp c$ the hyperfine structure remains resolved throughout.
			Inset: low-field region for $H_0 \perp c$, showing orthorhombic
			sites present at every concentration (red bars), \ce{Dy^3+}-\ce{Dy^3+}
			pair centers emerging with increasing doping (blue bars), and the
			residual \ce{Er^3+} line identified from its angular dependence
			(purple bar, Fig.~\ref{fig:WFPG2}), visible in (s1) and (s2) only.}
		\label{fig:conc}
	\end{figure}
	
	Figure~\ref{fig:conc} shows the \ce{Dy^3+} EPR spectra recorded at 7~K
	with increasing dopant concentration, for $H_0 \perp c$ (panel a) and
	$H_0 \parallel c$ (panel b); the $H_0 \perp c$ spectra were recorded at
	the azimuth $\varphi \approx 24^{\circ}$ from $a$, where the in-plane
	linewidth is minimal (Fig.~\ref{fig:WFPG2}). In both panels, the blue,
	orange, and green traces correspond to samples (s1), (s2), and (s3),
	with EPR-calibrated concentrations of 80~ppb, 2.3~ppm, and 247~ppm,
	respectively (Table~\ref{tab:samples}). The linewidth increases
	steadily with concentration in both orientations. So does the
	lineshape asymmetry, although the effect is considerably more
	pronounced for $H_0 \parallel c$; for $H_0 \perp c$ it remains weaker,
	yet still clearly visible.
	
	The visibility of the \ce{^{161}Dy} and \ce{^{163}Dy} hyperfine
	satellites also depends on orientation. For $H_0 \parallel c$, they
	are no longer resolved in (s3): the linewidth at this concentration
	is large enough to blend them into a single broad envelope. For
	$H_0 \perp c$, by contrast, the hyperfine structure remains resolved
	up to (s3), a direct benefit of recording at the linewidth-minimizing
	azimuth.
	
	At low field, the spectra reveal several additional resonances,
	illustrated in the inset of panel (a) for $H_0 \perp c$ only. One set
	of lines (red bars) is present at every concentration, including in
	the nominally undoped sample (s1); we assign these to orthorhombic
	sites associated with charge compensation. A second set (blue bars)
	is absent in (s1), only weakly visible in (s2), and prominent in
	(s3); their growth with doping is consistent with
	\ce{Dy^3+}-\ce{Dy^3+} pair centers, whose statistical weight scales
	with the square of the dopant concentration. A further, isolated
	line (purple bar), visible in (s1) and (s2) but absent in (s3),
	corresponds to residual \ce{Er^3+}, unambiguously identified from the
	angular dependence of its resonance field (Fig.~\ref{fig:WFPG2}).
	
	\subsection{Internal electric fields}
	\label{sec:LEFE}
	
	The angular modulation of the Dy$^{3+}$ linewidth in the $ab$ plane, with its
	minimum near $\varphi \approx 24^{\circ}$, reflects the linear electric field
	effect (LEFE), first observed by Royce and Bloembergen for \ce{Cr^3+}:\ce{Al2O3}
	\cite{royce1963} and characterized in \Ca by Mims for Ce$^{3+}$ and Er$^{3+}$
	\cite{mims1965,mims1966}. In $S_4$ point symmetry, electric fields lying in
	the $ab$ plane produce no shift of the resonance frequency when $H_0$ also
	lies in that plane; only a field component along $c$ displaces $g_\perp$
	linearly and shifts the resonance field~\cite{mims1965}. Charge-compensating
	point defects, distributed at positions uncorrelated with each paramagnetic
	dopant, generate a random distribution of such $c$-directed internal fields,
	consistent with the local-field picture developed by Mims and Gillen for
	charge-compensated sites in \Ca \cite{mims1967}.
	The resulting spread in resonance conditions broadens the line with an
	angular profile that follows the single-ion LEFE sensitivity: the linewidth
	is maximum where $\partial g_\perp^2 / \partial E_c$ is largest, and it
	vanishes at the angle $\varphi_0$ where that sensitivity is zero
	\cite{mims1966}. Quantitative LEFE coefficients have previously been reported
	for Yb$^{3+}$ and Mn$^{2+}$ in scheelite hosts \cite{kiel1970,kiel1971} and for
	Gd$^{3+}$ in \Ca and \ce{SrMoO4} under an externally applied field
	\cite{nepsha1969}, but, to our knowledge, no quantitative characterization of
	this effect has previously been reported for Dy$^{3+}$ in any host lattice.
	
	The LEFE coupling coefficients of Dy$^{3+}$ in \Ca are unknown a priori, so
	the linewidth modulation of Dy$^{3+}$ alone cannot be converted to an
	internal field amplitude. We circumvented this by using the residual
	Er$^{3+}$ impurity present in samples (s1) and (s2) as an in-situ
	calibration probe, recording its linewidth over the identical $ab$-plane
	sweep. The Er$^{3+}$ linewidth is well described by
	
	\begin{equation}
		\Gamma(\varphi) = \Gamma_0 + \Gamma_\varphi
		\left|\sin\!\left[2(\varphi-\varphi_0)\right]\right|,
		\label{eq:lefefit}
	\end{equation}
	
	\noindent the functional form predicted by Mims' theory for a random
	distribution of point charges at an $S_4$ site~\cite{mims1966}. The fits are
	shown in Figs.~\ref{fig:LWErPG2} and~\ref{fig:LWEr10ppm} of
	Appendix~\ref{app:LEFE}.
	For sample (s1), the parameters are $\Gamma_0 = 9.0 \pm 0.5$~MHz,
	$\Gamma_\varphi = 50\pm 2$~MHz, and $\varphi_0 = 30.8^{\circ}$; for sample
	(s2), $\Gamma_0 = 19.5\pm 0.9$~MHz, $\Gamma_\varphi = 155 \pm 8$~MHz, and
	$\varphi_0 = 31.4^{\circ}$. The phase $\varphi_0 \approx 31^{\circ}$ is
	consistent between the two samples, matches the value tabulated by Mims for
	Er$^{3+}$ in \Ca \cite{mims1965}, and agrees with the independent
	measurement of Le~Dantec\etal on a separately grown crystal
	\cite{ledantec2021}. Such cross-sample reproducibility confirms that
	$\varphi_0$ is set by the LEFE tensor of Er$^{3+}$ at the $S_4$ site,
	independent of dopant concentration or crystal provenance. Using the
	coupling coefficient $\alpha_\mathrm{Er} = 11 \times 10^{-6}$~cm/V reported
	by Mims for Er$^{3+}$ in \Ca \cite{mims1965}, $\Gamma_\varphi$ converts to
	internal fields of $E_\mathrm{int} = 68$~kV/cm for sample (s1) and
	$E_\mathrm{int} = 209$~kV/cm for sample (s2). Sample (s3) contains no
	detectable Er$^{3+}$ impurity; it cannot be characterized by this approach.
	
	\begin{figure}[h]
		\includegraphics[width=\columnwidth]{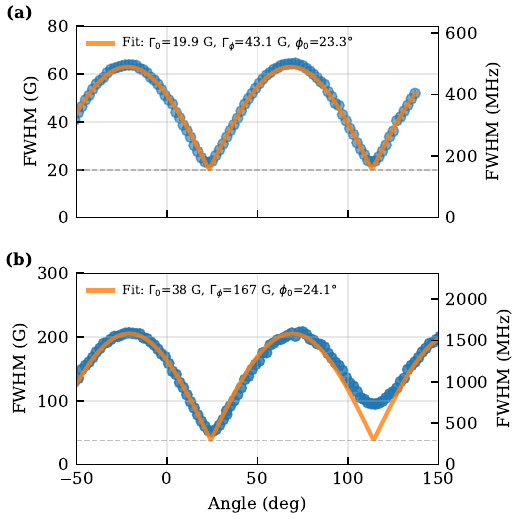}
		\caption{Angular dependence of the Dy$^{3+}$ linewidth (FWHM) for $H_0$
			rotated in the $ab$ plane, in field units (left axis) and in frequency
			units using $g_\perp(\mathrm{Dy^{3+}}) = 5.45$ (right axis).
			(a) Sample (s1). The solid line is a fit to Eq.~(\ref{eq:lefefit}),
			with $\Gamma_0 = 152\pm 5$~MHz, $\Gamma_\varphi = 329 \pm 10$~MHz, and
			$\varphi_0 = 23.3^\circ$. (b) Sample (s2); best-fit parameters:
			$\Gamma_0 = 290\pm 15$~MHz, $\Gamma_\varphi = 1274\pm 61$~MHz,
			$\varphi_0 = 24.1^\circ$. Dashed lines indicate $\Gamma_0$.}
		\label{fig:LWDy}
	\end{figure}
	
	The Dy$^{3+}$ linewidth modulation follows the same form,
	Eq.~(\ref{eq:lefefit}), with the fits shown in Fig.~\ref{fig:LWDy}. The
	parameters are $\Gamma_0 = 152\pm 5$~MHz, $\Gamma_\varphi = 329\pm 10$~MHz,
	$\varphi_0 = 23.3^{\circ}$ for sample (s1), and $\Gamma_0 = 290\pm 15$~MHz,
	$\Gamma_\varphi = 1274\pm 61$~MHz, $\varphi_0 = 24.1^{\circ}$ for sample (s2).
	The phase $\varphi_0 \approx 24^{\circ}$ is stable between the two samples
	and distinct from the $31^{\circ}$ observed for Er$^{3+}$: the two ions
	carry different LEFE tensors at the same site.
	
	Combining the Er$^{3+}$-calibrated internal fields with the Dy$^{3+}$
	angular data, and using the relation derived by Mims and Mashur for
	electric-field-induced $g$- and hyperfine shifts of rare-earth ions in
	scheelite hosts \cite{mims1972},
	\begin{equation}
		\alpha_\mathrm{Dy} = (B_{31}^2 + B_{36}^2)^{1/2} =
		\frac{2\,g_\perp\,h\,\Gamma_\varphi}{\mu_B H_0\,E_\mathrm{int}},
		\label{eq:alpha}
	\end{equation}
	where $H_0 = h\nu/(g_\perp \mu_B)$ is the resonance field of the ion
	considered and $\Gamma_\varphi$ is expressed in frequency units (the same
	relation, applied to Er$^{3+}$ with $\alpha_\mathrm{Er}$, underlies the
	internal-field calibration described above), we obtain
	$\alpha_\mathrm{Dy} = 30.6\times 10^{-6}$~cm/V from sample (s1) and
	$38.6\times 10^{-6}$~cm/V from sample (s2). Each determination combines two
	measured amplitudes, $\Gamma_\varphi(\mathrm{Dy})$ and, through
	$E_\mathrm{int}$, $\Gamma_\varphi(\mathrm{Er})$. Assigning a 10\% relative
	uncertainty to each amplitude (see Figs.~\ref{fig:LWErPG2}
	and~\ref{fig:LWEr10ppm}) and adding the two contributions in
	quadrature yields 14\% per determination. On this budget the two values
	differ by only $1.15\,\sigma$; they are mutually consistent. The weighted
	mean is
	\begin{equation}
		\alpha_\mathrm{Dy} = (34 \pm 5)\times 10^{-6}~\text{cm/V},
		\label{eq:alphaDy}
	\end{equation}
	where the quoted uncertainty includes both fit uncertainties and the 10\%
	accuracy of $\alpha_\mathrm{Er}$ in the literature.
	From $\tan(2\varphi_3) = -B_{31}/B_{36}$, the individual tensor components
	evaluated at $\varphi_0 = 23.7^{\circ}$ \add{(the mean of the two
		determinations obtained with an \ce{Er^3+} calibration; sample (s3), which
		has none, gives $22.4^{\circ}$)} are
	\begin{equation}
		B_{36} = (+23\pm 4)\times 10^{-6}~\text{cm/V}, \quad
		B_{31} = (-25\pm 4)\times 10^{-6}~\text{cm/V}.
		\label{eq:B31B36}
	\end{equation}
	The $\pm2^{\circ}$ alignment uncertainty on $\varphi_0$ propagates to about
	7\% on each component, small compared with the 14\% amplitude budget. The
	near-equality $|B_{31}| \approx |B_{36}|$ is reflected in
	$\varphi_3 \approx 22.5^{\circ}$ (the value obtained when the two magnitudes
	are exactly equal), consistent with the fitted $23$--$24^{\circ}$. For
	Er$^{3+}$, by contrast, $|B_{31}|$ is roughly twice $|B_{36}|$, shifting
	$\varphi_0$ to $31^{\circ}$; the two ions differ in their LEFE tensor
	structure, as expected from their distinct $4f$ wavefunctions at the $S_4$
	site. To our knowledge these are the first LEFE parameters reported for
	Dy$^{3+}$ in CaWO$_4$.

	\subsection{Second-order electric field effect and lineshape asymmetry}
	\label{sec:LEFE2}
	
	We now return to the lineshape asymmetry noted in
	Sec.~\ref{sec:epr-params} and left unexplained there. The first-order
	LEFE characterized above cannot produce it. The two inversion-related
	\ce{Ca^2+} sites of the scheelite cell experience first-order shifts of
	opposite sign, $+\delta$ and $-\delta$, for the same local field;
	averaged over a symmetric distribution of internal fields, these shifts
	broaden the line but leave it symmetric. Any first-order mechanism is
	bound by this cancellation. The second-order Stark effect is not.
	
	The second-order correction to the energy of the ground doublet reads
	\begin{equation}
		\Delta\varepsilon^{(2)} =
		\sum_{n\neq i}
		\frac{\left|\braket{n|\mathcal{H}'_E|\psi_i}\right|^2}
		{E_i^{(0)}-E_n^{(0)}},
		\label{eq:secondorder}
	\end{equation}
	where $\mathcal{H}'_E$ is the electric perturbation of the crystal
	field and the sum runs over the excited states of the $^{6}H_{15/2}$
	manifold. Three properties of Eq.~\eqref{eq:secondorder} govern the
	phenomenology. The shift is quadratic in $E$. Because $(-E)^2 = E^2$,
	the two inversion-related sites now shift in the \emph{same}
	direction, so the ensemble cancellation that protects the first order
	is absent. And since the numerator is a squared modulus, the sign of
	the shift is fixed by the energy denominators, not by the orientation
	of the local field. A symmetric distribution of internal fields is
	thus mapped onto a one-sided distribution of resonance shifts: the
	line acquires a tail on one side only, which is precisely an
	asymmetric profile.
	
	The orientation dependence supports this assignment. In $S_4$
	symmetry, no component of $\mathbf{E}$ transforms as the totally
	symmetric representation, so the first-order diagonal matrix elements
	vanish identically for $H_0 \parallel c$; in the $ab$ plane, the
	first-order sensitivity crosses zero at the azimuth $\varphi_0$. The
	second-order term survives both suppressions, since it relies on
	off-diagonal matrix elements whose squared moduli never vanish
	simultaneously. Two observations follow. The asymmetry should be most
	visible where the first order is silent, i.e., for $H_0 \parallel c$,
	and this is indeed the orientation where Fig.~\ref{fig:conc} shows
	the strongest Dysonian character. It should also persist at
	$\varphi \approx \varphi_0$ in the $ab$ plane, where the first-order
	broadening reaches its floor $\Gamma_0$.
	
	The case for a second-order contribution rests, first, on the
	magnitude of that floor itself. At $\varphi = \varphi_0$ with
	$H_0 \perp c$, the first-order LEFE broadening is nulled by
	construction, and the width should fall back to whatever
	concentration-dependent mechanisms remain. It does not. The floors
	extracted from Eq.~\eqref{eq:lefefit} are $\Gamma_0 = 19.9$,
	$38$ and $202$~G for (s1), (s2) and (s3) (i.e., 152, 290 and
	1540~MHz with $g_\perp = 5.45$; Figs.~\ref{fig:LWDy}
	and~\ref{fig:LWDy2000ppm}), values that are anomalously large for a
	dilute rare-earth system. By comparison, the Er$^{3+}$ probe in the
	very same crystal shows a floor of only $0.77$~G ($9.0$~MHz) in (s1).
	Dipolar broadening cannot account for this.
	For a random three-dimensional distribution of like spins, the
	Portis-Castner expression gives a full width
	\begin{equation}
		\Gamma_\mathrm{dip} = \frac{\mu_0 (g\mu_B)^2 C}{9\sqrt{3}\,h},
		\label{eq:dipolar}
	\end{equation}
	where $C$ is the spin density \cite{portis1953,castner1959,abragam2013}.
	Evaluated at the EPR-calibrated concentrations of
	Table~\ref{tab:samples} with $g_\perp = 5.45$, it yields
	$4\times10^{-5}$, $1.2\times10^{-3}$ and $0.13$~G for (s1), (s2) and
	(s3). Even in the most concentrated sample, the dipolar width falls
	three orders of magnitude short of the measured floor. The dopant
	concentration must therefore act on the linewidth through another
	channel.
	
	The LEFE coupling itself points to that channel. The coefficient
	measured in Sec.~\ref{sec:LEFE},
	$\alpha_\mathrm{Dy} = (34\pm5)\times10^{-6}$~cm/V, is three times
	$\alpha_\mathrm{Er} = 11\times10^{-6}$~cm/V, until now the largest
	linear electric coupling reported for a rare-earth ion in this host
	\cite{mims1965}. An ion this susceptible to electric fields at first
	order can be expected to respond at second order as well. The known
	$\alpha_\mathrm{Dy}$ also lifts the limitation encountered in
	Sec.~\ref{sec:LEFE} for sample (s3): although this crystal contains
	no Er$^{3+}$ probe, its internal field is now obtained directly from
	the Dy$^{3+}$ angular modulation,
	$\Gamma_\varphi = 2501$~MHz ($328$~G, Fig.~\ref{fig:LWDy2000ppm}), by
	inverting Eq.~\eqref{eq:alpha}, which gives
	$E_\mathrm{int} = 466$~kV/cm. We note in passing that the
	internal fields, 68, 209 and 466~kV/cm, do not track the Dy$^{3+}$
	content, which spans more than three orders of magnitude across the
	three samples; the density of compensating defects is evidently set
	by the growth conditions of each boule rather than by the dopant
	level alone.
	
	Having the electric fields and linewidth floors in hand for all three samples, the
	second-order scaling can be tested. A shift quadratic in $E$ predicts
	a broadening proportional to $\langle E^2 \rangle$. Writing
	\begin{equation}
		\Gamma_0(E_\mathrm{int}) = \Gamma_{00} + \beta E_\mathrm{int}^2,
		\label{eq:floor}
	\end{equation}
	the best fit gives $\Gamma_{00} = 15.2$~G and
	$\beta = 7.2\times10^{-4}$~G\,(kV/cm)$^{-2}$ (see
	Fig.~\ref{fig:LEFE2nd} \add{of Appendix~\ref{app:LEFE-theory}}). Three samples
	provide only three points, but they span nearly a decade in
	$E_\mathrm{int}$ and follow the quadratic dependence expected for a
	second-order electric field effect.
	\add{A linear dependence on $E_\mathrm{int}$ is excluded by the same three
		points. Anchored on (s1) and (s3), it would place the floor of (s2) near
		$85$~G against the measured $38$~G, and would require a negative intercept
		of about $-11$~G. The quadratic form needs neither.}
	We also note that $E_\mathrm{int}$
	for (s3) derives from the first-order modulation amplitude
	$\Gamma_\varphi$, a fit parameter independent of the floor $\Gamma_0$
	against which it is compared.
	\add{The residual scatter about Eq.~\eqref{eq:floor} nevertheless reaches
		some $20\%$ on (s2) and (s3). Since $\Gamma_{00}$ collects whatever is
		independent of the internal field, and the three crystals come from three
		independent boules with different co-doping chemistry, a single common
		$\Gamma_{00}$ is an idealization; the scatter is of the size that
		boule-to-boule variation would produce.}
	The origin of $\Gamma_{00}$, and the
	reason it exceeds the corresponding Er$^{3+}$ floor by more than an
	order of magnitude in the same crystal, are left open at this stage.
	\add{We note only that the ground doublet of \ce{Dy^3+} sits
		\textcolor{black}{$3.4~\mathrm{cm}^{-1}$} below a partner of the same symmetry
		(Sec.~\ref{sec:reconciling}), so its $g$ tensor is first-order sensitive to
		any perturbation of the tetragonal field, strain included, and not to
		electric fields alone. Whether this accounts for $\Gamma_{00}$ would
		require a strain-dependent measurement that we have not performed.}
	
	The same term closes the argument opened at the start of this
	section. A quadratic shift displaces every ion to the same side of
	the unperturbed resonance, by an amount distributed as
	$E_\mathrm{int}^2$; the line therefore develops a one-sided tail
	whose weight grows with $\langle E_\mathrm{int}^2 \rangle$, and the
	Dysonian asymmetry of Fig.~\ref{fig:conc} strengthens with doping
	exactly as this picture requires.

	\subsection{Microscopic origin of the electric-field coupling strength}
	\label{sec:LEFE-origin}
	
	The coupling reported here does not simply extend the ranking of the
	scheelite series; it splits it in two. Among the Kramers ions studied by
	the electron-echo method in \Ca, the amplitude
	$\alpha \equiv (B_{31}^2+B_{36}^2)^{1/2}$ is
	$8.7\times10^{-8}$, $3.7\times10^{-8}$, and $8.1\times10^{-8}~\mathrm{cm/V}$
	for \ce{Ce^3+}, \ce{Nd^3+}, and \ce{Yb^3+}, respectively, but
	$1.1\times10^{-5}~\mathrm{cm/V}$ for \ce{Er^3+}~\cite{mims1965}, a jump of
	more than two orders of magnitude. The coefficient obtained in
	Sec.~\ref{sec:LEFE},
	$\alpha_\mathrm{Dy} = (34\pm5)\times10^{-6}~\mathrm{cm/V}$, exceeds even
	this erbium value, by a factor of $\approx3$. Dysprosium and erbium
	therefore occupy a regime of their own, separated from each other by less
	than one order of magnitude while both sit two to three orders above
	cerium, neodymium, and ytterbium. Any microscopic account has to explain
	the cluster, not just the ranking.
	
	Appendix~\ref{app:LEFE-symmetry} shows that the field-induced increments
	$\delta B_q^k$ active for $\mathbf{E}\parallel c$ carry $k=\pm2$, and that
	these operators connect the $\Gamma_{5,6}$ ground manifold exclusively to
	$\Gamma_{7,8}$ states. The first-order shift is therefore a cross term
	between $\delta\mathcal{H}_E$ and the Zeeman interaction, mediated by the
	nearest doublet of the opposite representation,
	\begin{equation}
		\delta g \propto \sum_{n\in\Gamma_{7,8}}
		\frac{\langle 0|\delta\mathcal{H}_E|n\rangle
			\langle n|\mu_B\,\mathbf{B}_0\!\cdot\!g_J\mathbf{J}|0'\rangle}
		{\Delta_n},
		\label{eq:LEFEcross}
	\end{equation}
	and the same energy denominator controls the second-order shift of
	Sec.~\ref{sec:LEFE2}, which is not subject to the $S_4$ selection
	rule and survives under every field orientation. The relevant gap is thus
	to the first \emph{opposite-symmetry} doublet, not to the nearest doublet
	of any symmetry.
	
	This is not an academic distinction for \ce{Dy^3+}. 	\add{The same distinction has to be applied to the comparison ions, and
		\ce{Er^3+} is the one that matters, since it carries the only other
		coupling in the same decade. Its ground doublet has crystal quantum
		number $\mu = 3/2$, with amplitudes on $|{\pm}13/2\rangle$,
		$|{\pm}5/2\rangle$, $|{\pm}3/2\rangle$ and $|{\pm}11/2\rangle$ and
		strictly none on $|{\pm}1/2\rangle$~\cite{bernal1971}: erbium and
		dysprosium share the same $\Gamma_{5,6}$ ground representation. The
		first $\Gamma_{7,8}$ level of the $^4I_{15/2}$ manifold lies at
		$19.2~\mathrm{cm^{-1}}$, the second at $51.1~\mathrm{cm^{-1}}$, with a
		$\Gamma_{5,6}$ doublet at $24.9~\mathrm{cm^{-1}}$ in
		between~\cite{bernal1971}. The lowest of these is the doublet whose EPR
		Antipin \etal resolved above 8~K, $g_\parallel = 3.42$ and
		$g_\perp = 6.98$ against 3.65 and 7.03
		calculated~\cite{antipin1968,bernal1971}, and it is also the level their
		Orbach analysis of $T_1$ returns, $\Delta_1 = 18~\mathrm{cm^{-1}}$.
		Optical energies, excited-state $g$ tensor and relaxation data converge
		on one assignment.}
	
	\add{Two consequences follow. The gap that governs Eq.~\eqref{eq:LEFEcross}
		is the same for both ions to within \textcolor{black}{twenty} percent, $19~\mathrm{cm^{-1}}$
		for erbium against \textcolor{black}{$16$--$18~\mathrm{cm^{-1}}$} for dysprosium, whereas
		the ground doublets of \ce{Nd^3+}, \ce{Yb^3+} and \ce{Ce^3+} are isolated
		by $\gtrsim 100~\mathrm{cm^{-1}}$; values from the compilation in
		Ref.~\cite{mims1967}. The bimodal split of the series therefore does
		track $1/\Delta_1$, and it is not fortuitous that the two $J=15/2$ ions
		dominate: eight Kramers doublets compressed into roughly
		$400~\mathrm{cm^{-1}}$ almost guarantee a low-lying neighbor of the
		required symmetry, something the isolated ground doublets of the other
		three do not offer. But the same equality of the denominators means that
		$1/\Delta_1$ cannot account for the factor of three separating the two
		members of the cluster.}
	
	\add{That factor is carried by the numerator, and the rank-2 Stevens
		coefficient supplies it. For \ce{Dy^3+}, $\alpha_J = -2/315 =
		-6.35\times10^{-3}$; for \ce{Er^3+}, $\alpha_J = +4/1575 =
		+2.54\times10^{-3}$, a ratio of 2.5 in magnitude. Combined with the
		measured gaps this predicts $\alpha_\mathrm{Dy}/\alpha_\mathrm{Er}$
		between \textcolor{black}{2.7 and 2.9}, according to whether the fitted or the optical value
		is taken for $\Delta_1(\mathrm{Dy})$, against $3.1 \pm 0.6$ observed. The
		estimate holds the odd-parity coupling coefficients $R^{(r)}_{02}$ of
		Appendix~\ref{app:LEFE-symmetry} equal for the two ions, which they need
		not be, so a residual of a few tens of percent is the expected size of
		that assumption rather than a discrepancy.}
	
	\add{The free ion cannot carry the argument on its own, as \ce{Ce^3+}
		shows: it has the largest $\alpha_J$ of the series
		($-5.71\times10^{-2}$) together with one of the smallest couplings, no
		Stevens factor compensating a denominator an order of magnitude larger.}
	The ionic-displacement mechanism of Kiel and Mims~\cite{kiel1970}, which
	correctly accounts for the growth of the shifts from \Ca\ to \ce{SrWO4}
	and \ce{BaWO4}, predicts the largest effect for the loosest ion in the
	\ce{Ca^2+} site, namely \ce{Yb^3+}, whose coupling is among the smallest.
	A dominant role for the $4f$--$5d$ electronic channel would favor
	\ce{Ce^3+}, for which this gap is minimal across the series. \add{Both
		counterexamples point back to the crystal-field level structure as the
		quantity that sets the scale, the free-ion Stevens factor ordering the
		ions only within a group sharing a comparable $\Delta_1$.}
	
	\section{Conclusion}
	\label{sec:conclusion}
	
	We have established the EPR spectroscopy of \ce{Dy^3+} at the $S_4$ site of
	\Ca across more than three decades of concentration. The $g$ factors
	and the \ce{^{161}Dy}/\ce{^{163}Dy} hyperfine constants refine the 1966
	determination of Antipin \etal by an order of magnitude, and the
	ratio $A_\parallel g_\perp / A_\perp g_\parallel = 0.973 \pm 0.020$ bounds the
	$J$-mixing of the ground doublet at the percent level. A joint fit of the $S_4$ crystal-field parameters, first within the
	${}^6H_{15/2}$ multiplet and then within the complete $^6H$ term, to
	the optical levels of Wortman and Sanders and to the measured $g$
	tensor resolves the long-standing disagreement between optical and EPR
	determinations: the optical energies alone leave the composition of
	the ground doublet, and with it the $g$ tensor, essentially
	unconstrained, and a parameter set fixed by the optical data yields
	$g$ values far from experiment. The joint fit removes this ambiguity
	without degrading the optical residual, and identifies the ground
	doublet as an admixture of \textcolor{black}{$|13/2\rangle$ and $|{-11/2}\rangle$}
	character within $\Gamma_{5,6}$. \add{Tested on the excited multiplets of
		the same term, which entered no fit, it reproduces the seven
		${}^6H_{13/2}$ crystal-field levels better than the optical parameters
		themselves.} It places a second doublet of the
	same symmetry at $2$--$5~\mathrm{cm}^{-1}$ above the ground state, with
	\textcolor{black}{$g_z \approx 5.1$ and $g_\perp \approx 6.4$}, a prediction open to test
	by far-infrared or high-frequency EPR spectroscopy.
	
	The linewidth carries the signature of the electrostatic environment. Its
	$ab$-plane modulation, calibrated in situ against the \ce{Er^3+} impurity
	measured over the identical angular sweep, gives the first linear
	electric-field parameters for \ce{Dy^3+} in any host,
	$\alpha_\mathrm{Dy} = (34 \pm 5)\times10^{-6}$~cm/V with
	$|B_{31}| \approx |B_{36}|$. The anomalously large angular floor of the
	linewidth, its quadratic scaling with the internal field, and the one-sided
	lineshape asymmetry all point to a second-order Stark shift, which no $S_4$
	selection rule suppresses. The strength of both orders follows from the
	\textcolor{black}{16--18}~cm$^{-1}$ gap to the first $\Gamma_{7,8}$ doublet; it is the
	crystal-field level structure of the $J=15/2$ ion, not any free-ion
	property, that controls the coupling. \add{The same term sets a practical
		limit on the spectroscopy itself: because the quadratic shift is largest
		where the linear one is silent, namely for $H_0 \parallel c$, it is the
		precision attainable on $g_\parallel$ that it degrades, which is where the
		present determination and the earlier tabulated value disagree.}
	
	These results carry practical weight for the scheelite quantum platform. An
	electric coupling three times that of \ce{Er^3+} makes \ce{Dy^3+} the most
	field-sensitive rare-earth spin in \Ca. In practical terms,
	Eq.~(\ref{eq:alpha}) can be read as a tunability,
	$\delta\nu/\nu = \alpha_\mathrm{Dy} E / (2 g_\perp^2)$: at the azimuth of
	maximum sensitivity, a field of 1~kV/cm applied along $c$ shifts an
	individual spin by 5.4~MHz at X band, and separates the two
	inversion-related sublattices by twice that. This is the figure of merit
	relevant to the electrical gating of individual spins, and equally the
	figure that sets the sensitivity to charge noise from compensating defects,
	which will dominate decoherence at $S_4$ sites unless growth conditions
	suppress them. The internal fields measured here, 68 to 466~kV/cm in boules
	of different provenance, quantify that environment and provide the input
	needed to assess \ce{Dy^3+} as a candidate for electrically addressable spin
	qubits.
	
	\begin{acknowledgments}
		Financial support from the IR INFRANALYTICS FR2054 for conducting the research is gratefully
		acknowledged. This work was funded, in part, by the French National Research Agency (ANR) under project “ANR-24-CE29-6059-01”, "ANR-22-PETQ-0010"(Qmemo), ANR-24-CE47-1190 (CHORIZO) and PEPR MolQif.  A.M.K. is supported by ANR QuantEdu-France (22-CMAS0001) and by France 2030 investment plan, as part of the Initiative d’Excellence d’Aix-Marseille Université - SpinEClock project A*MIDEX (AMX-22-RE-AB-199) .
	\end{acknowledgments}
	
	\bibliography{CaWO4Dy_final}

	\clearpage
	\onecolumngrid
	
	\appendix
	
	\section{Crystal-field analysis of the ground doublet}
	\label{app:CF}
	
	This appendix documents the crystal-field calculation behind
	Sec.~\ref{sec:reconciling}: the model and its two levels of
	approximation, the construction of the matrix elements, the fitting
	procedure, \add{the test of the result on the excited multiplets,} and the
	comparison between the intermediate and final results.
	
	\subsection{Model}
	
	The Hamiltonian is
	\begin{equation}
		H = H_{\mathrm{FI}} + \sum_{k=2,4,6}\sum_{q} B_{kq}\, C^{(k)}_q ,
		\label{eq:HCF}
	\end{equation}
	where $H_{\mathrm{FI}}$ is the free-ion Hamiltonian (electrostatic
	repulsion and spin-orbit coupling), whose eigenstates within the
	$4f^9$ configuration are the $LS$ terms and, once spin-orbit coupling
	is included, their $J$ multiplets; ${}^6H_{15/2}$ is the lowest such
	multiplet. In the Wybourne convention adopted here, the $C^{(k)}_q$
	are the true spherical-tensor crystal-field operators (spherical
	harmonics acting on the orbital coordinates) summed over the $4f$
	electrons, not to be confused with the effective-spin operators of
	Eq.~(\ref{eq:spinham}): the diagonalization is carried out in
	the multielectron basis, prior to any spin-Hamiltonian reduction. At
	the $S_4$ site only $B_{20}$, $B_{40}$, $B_{60}$, $B_{44}$ and
	$B_{64}$ are nonzero, and the in-plane axes are fixed so that $B_{44}$
	is real.
	
	Two truncations of the basis are used in sequence. The
	\emph{intra-multiplet} model diagonalizes Eq.~(\ref{eq:HCF}) within
	${}^6H_{15/2}$ alone (16 states); the Zeeman interaction then reduces
	to $g_J \mu_B \mathbf{H}_0 \cdot \mathbf{J}$ with $g_J = 4/3$. The
	\emph{full-term} model diagonalizes it within the complete $^6H$ term
	($J = 15/2, 13/2, \ldots, 5/2$, 66 states), with the multiplet
	barycenters set to their measured values and the Zeeman operator taken
	as $\mu_B \mathbf{H}_0 \cdot (\mathbf{L} + 2\mathbf{S})$, including
	the matrix elements connecting different $J$ multiplets. The second
	model therefore treats exactly the $J$-mixing that the crystal field
	induces, which the first can only absorb into effective parameter
	values.
	
	Three approximations bound the accuracy of the full-term model. We
	work in pure $LS$ coupling within $^6H$, which is justified a
	posteriori: the fitted ground doublet is $99.96\%$ ${}^6H_{15/2}$, and
	the measured ratio $A_\parallel g_\perp / A_\perp g_\parallel =
	0.973 \pm 0.020$, which would be unity for a single-$J$ doublet, limits any
	term mixing to the percent level. The $^6F$ multiplets, lying above
	$7500~\mathrm{cm}^{-1}$ \cite{wortman1971}, are excluded. Only the
	static crystal field is retained; vibrational corrections to $g$ are
	neglected.
	
	\subsection{Matrix elements}
	
	The reduced matrix elements $\langle {}^6H, L{=}5 \| \sum_i C^{(k)}(i)
	\| {}^6H, L{=}5\rangle$ are computed exactly from the stretched state
	$|M_L{=}5, M_S{=}5/2\rangle$ of $4f^9$, which is a single Slater
	determinant: seven spin-up electrons fill the $m_l$ subshell and
	contribute nothing for $k>0$, leaving only the two spin-down electrons
	in $m_l = 3$ and $2$. The expectation value of $C^{(k)}_0$ in this
	state is evaluated one electron at a time and the Wigner--Eckart
	theorem is then inverted. No tabulated Stevens factors enter the
	calculation; as a check, the $k=2$ element reproduces the tabulated
	$\alpha_J = -2/315$ of Dy$^{3+}$ \add{once the Wybourne-to-Stevens
		factor $\lambda_{20} = 1/2$ is applied, i.e.\ $\langle 15/2,15/2 |
		\sum_i C^{(2)}_0 | 15/2,15/2\rangle = -1/3$}. Reduced elements between different
	$J$ multiplets, $\langle LSJ \| C^{(k)} \| LSJ'\rangle$, follow from
	standard $6j$ recoupling, the $C^{(k)}$ acting on the orbital space
	only, and the reduced elements of $\mathbf{L} + 2\mathbf{S}$ are built
	in the same way. As an internal control, the diagonal element $\langle
	15/2, 15/2 | L_z + 2S_z | 15/2, 15/2 \rangle = g_J J = 10$ is
	recovered to machine precision.
	
	Under $S_4$ the projection $M$ is conserved modulo 4, so $B_{44}$ and
	$B_{64}$ couple states only in steps $\Delta M = \pm 4$; the Kramers
	doublets separate into the two irreducible pairs $\Gamma_{5,6}$ and
	$\Gamma_{7,8}$ according to $M \bmod 4$. For a Kramers doublet
	$(\psi_+, \psi_- = \Theta \psi_+)$, with $\Theta$ the time-reversal
	operator, the principal $g$ values are
	\begin{equation}
		g_z = 2\,\langle \psi_+ | L_z + 2S_z | \psi_+ \rangle, \qquad
		g_\perp = 2\,\bigl| \langle \psi_+ | L_x + 2S_x | \psi_- \rangle \bigr|,
		\label{eq:gdef}
	\end{equation}
	which reduce, within one $J$ multiplet, to $g_z = 2 g_J \langle J_z
	\rangle$ and $g_\perp = 2 g_J |\langle \psi_+ | J_x | \psi_-
	\rangle|$. In practice the degenerate Kramers partners returned by the
	numerical diagonalization are first disentangled by diagonalizing
	$J_z$ (or $L_z + 2S_z$) in the two-dimensional degenerate subspace,
	and the full $g$ tensor is obtained from the $2\times 2$ matrices
	$A_a$ of the three magnetic-moment components within the doublet,
	through $K_{ab} = \tfrac12 \mathrm{Re}\,\mathrm{Tr}(A_a A_b^\dagger +
	A_b A_a^\dagger)$, whose eigenvalues give the squared principal $g$
	values.
	
	\subsection{Validation against Wortman and Sanders}
	
	Before any fitting, both models are evaluated at the nominal optical
	parameters of Ref.~\cite{wortman1971}, $(B_{20}, B_{40}, B_{60},
	B_{44}, \mathrm{Re}\,B_{64}) = (428, -825, -6.6, 972,
	448)~\mathrm{cm}^{-1}$. The intra-multiplet model gives $(g_\parallel,
	g_\perp) = (12.40, 2.60)$ for the ground doublet and $(4.55, 8.88)$
	for the first excited one, against $(-12.431, 2.494)$ and $(4.554,
	8.936)$ in Table~7 of Ref.~\cite{wortman1971} (cw EPR fixes only
	$|g|$); this validates the reduced matrix elements and conventions.
	The full-term model at the same parameters gives $(12.43, 2.41)$, with
	a $99.95\%$ ${}^6H_{15/2}$ weight: the shift relative to the 16-state
	values is a direct measure of the crystal-field-induced $J$-mixing,
	small on $g$ but, as shown below, not negligible for the level
	positions near $385~\mathrm{cm}^{-1}$.
	
	\add{A second and sharper control is available on the excited
		multiplets. Evaluated at the same optical parameters, the full-term model
		reproduces the ${}^6H_{13/2}$ and ${}^6H_{11/2}$ level schemes calculated
		in Ref.~\cite{wortman1971} to $0.32$ and $0.13~\mathrm{cm}^{-1}$ rms
		respectively, once the constant barycenter offset is removed. The two
		calculations therefore agree on the crystal-field physics to a fraction of
		a wavenumber, and any discrepancy in what follows reflects the parameter
		values alone.}
	
	\subsection{Two-step combined fit}
	
	The five parameters are adjusted to minimize
	\begin{equation}
		\chi^2 = \sum_{n=1}^{8} \frac{\bigl(E_n^{\mathrm{calc}} -
			E_n^{\mathrm{exp}}\bigr)^2}{\sigma_{E,n}^2}
		+ \frac{\bigl(g_\parallel^{\mathrm{calc}} - g_\parallel\bigr)^2}{\sigma_g^2}
		+ \frac{\bigl(g_\perp^{\mathrm{calc}} - g_\perp\bigr)^2}{\sigma_g^2},
		\label{eq:chi2}
	\end{equation}
	where the $E_n^{\mathrm{exp}}$ are the eight ${}^6H_{15/2}$ levels of
	Ref.~\cite{wortman1971} (0, 5, 18, 63, 84, 118, 381,
	$388~\mathrm{cm}^{-1}$), with uncertainties $\sigma_{E,n} =
	4$--$6~\mathrm{cm}^{-1}$ enlarged to 5 and $10~\mathrm{cm}^{-1}$ for
	the two tentative assignments at 5 and $84~\mathrm{cm}^{-1}$.
	\add{The two $g$ values are entered with a weight $\sigma_g = 0.005$,
		tighter than the experimental uncertainty of Eq.~(\ref{eq:fitvalues}), so
		that the fit is driven to reproduce them rather than to trade them against
		the optical residual; this is a weighting choice, not a claim about the
		measurement.} \textcolor{black}{Minimization uses the Nelder--Mead simplex,
		restarted from one hundred random points drawn in the five-dimensional
		parameter space. The $\chi^2$ landscape is rugged, above all in the
		$(B_{44}, B_{64})$ plane, and a single local descent converges to
		spurious minima that lie one order of magnitude above the best solution.}
	
	Step 1 restricts the model to the 16 states of ${}^6H_{15/2}$. The fit
	converges to \textcolor{black}{$(455, -906, -15, 821, 662)~\mathrm{cm}^{-1}$} and
	reproduces the measured $g$ values exactly (Table~\ref{tab:cf-fits}).
	\textcolor{black}{The optical residual is $2.5~\mathrm{cm}^{-1}$ rms, the two
		highest doublets being calculated at 377 and $391~\mathrm{cm}^{-1}$
		against the observed 381 and $388~\mathrm{cm}^{-1}$. A single-$J$ model
		is therefore already sufficient to reconcile the optical spectrum with
		the $g$ tensor, provided the parameter space is searched globally: the
		agreement does not depend on the $J$-mixing that a truncated basis
		excludes by construction. What the truncated basis does leave open is
		whether the parameters themselves are effective ones, absorbing in
		$B_{kq}$ what belongs to the inter-multiplet repulsion.}
	
	Step 2 \textcolor{black}{answers that question} by extending the basis to the full
	$^6H$ term, taking the step-1 parameters as the starting simplex. The
	fit converges to Eq.~(\ref{eq:Bkq-final})\textcolor{black}{, with $\chi^2 = 1.4$
		for ten observables}.
	Both $g$ values are again reproduced within $0.01$, and
	the optical rms \textcolor{black}{is $2.6~\mathrm{cm}^{-1}$, carried mainly by the
		two upper doublets, calculated at 377 and $392~\mathrm{cm}^{-1}$}.
	\textcolor{black}{The displacement from step 1 stays within a few percent on
		$B_{20}$, $B_{40}$ and $B_{44}$ and is nil on $\mathrm{Re}\,B_{64}$;
		only $B_{60}$ moves appreciably, from $-15$ to $+20~\mathrm{cm}^{-1}$,
		which is the $J$-mixing being reabsorbed into the parameter the data
		constrain least. The ground-doublet composition is identical to three
		decimals in the two models.} Comparing the three columns of
	Table~\ref{tab:cf-fits} makes the logic of the two-step procedure
	explicit. The intra-multiplet fit establishes that the $g$ tensor and
	the optical spectrum are compatible; the full-term
	fit, seeded by it, \textcolor{black}{confirms that the solution survives the
		exact treatment of $J$-mixing}, with residuals below the internal scatter of the
	optical data themselves.
	
	\begin{table}
		\caption{Crystal-field parameter sets and their predictions for
			Dy$^{3+}$:CaWO$_4$. First column: nominal optical parameters of
			Ref.~\cite{wortman1971} evaluated in the full-term (66-state) model.
			Second and third columns: the two steps of the combined fit to the
			eight optical levels and the measured $(g_\parallel, g_\perp) =
			\add{(7.22 \pm 0.01, 5.45 \pm 0.01)}$. Energies $E_1, E_2$ are the first two excited
			doublets; rms refers to the eight optical levels.}
		\label{tab:cf-fits}
		\begin{ruledtabular}
			\begin{tabular}{lccc}
				& W\&S \cite{wortman1971} & Step 1 & Step 2 \\
				& (optical only) & ($J=15/2$, 16 states) & (full $^6H$, 66 states) \\
				\colrule
				$B_{20}$ (cm$^{-1}$) & 428 & \textcolor{black}{455} & \textcolor{black}{449} \\
				$B_{40}$ (cm$^{-1}$) & $-825$ & \textcolor{black}{$-906$} & \textcolor{black}{$-919$} \\
				$B_{60}$ (cm$^{-1}$) & $-6.6$ & \textcolor{black}{$-15$} & \textcolor{black}{20} \\
				$B_{44}$ (cm$^{-1}$) & 972 & \textcolor{black}{821} & \textcolor{black}{812} \\
				$\mathrm{Re}\,B_{64}$ (cm$^{-1}$) & 448 & \textcolor{black}{662} & \textcolor{black}{662} \\
				\colrule
				$g_\parallel$ (calc) & 12.43 & \textcolor{black}{7.22} & \textcolor{black}{7.22} \\
				$g_\perp$ (calc) & 2.41 & \textcolor{black}{5.45} & \textcolor{black}{5.45} \\
				$E_1$ (cm$^{-1}$) & 4.0 & \textcolor{black}{4.2} & \textcolor{black}{3.4} \\
				$E_2$ (cm$^{-1}$) & 14.2 & \textcolor{black}{15.4} & \textcolor{black}{16.3} \\
				rms optical (cm$^{-1}$) & 9.0 & \textcolor{black}{2.5} & \textcolor{black}{2.6} \\
			\end{tabular}
		\end{ruledtabular}
	\end{table}
	
	\subsection{\add{Test on the excited multiplets}}
	\label{app:CF-multiplets}
	
	\add{Only the eight ${}^6H_{15/2}$ levels and the two measured $g$ values
		entered Eq.~(\ref{eq:chi2}). The five excited multiplets of the $^6H$ term
		nevertheless sit in the same 66-state basis, so their crystal-field
		splittings are predictions of the fitted $B_{kq}$, with no adjustable
		quantity left beyond the free-ion barycenter of each multiplet. Since the
		barycenters are spin-orbit quantities rather than crystal-field ones, each
		calculated multiplet is aligned on the observed barycenter and only the
		splitting pattern is compared. Wortman and Sanders located the seven
		Kramers doublets of ${}^6H_{13/2}$ by ${}^4F_{9/2}$ fluorescence
		\cite{wortman1971}, which makes that multiplet the cleanest test: it lies
		well below the $^6F$ term excluded from the basis, and every one of its
		levels is observed.}
	
	\add{Table~\ref{tab:multiplets} gives the comparison. The parameter set of
		Eq.~(\ref{eq:Bkq-final}) reproduces the seven ${}^6H_{13/2}$ levels to
		\textcolor{black}{$6.1~\mathrm{cm}^{-1}$} rms, against $9.8~\mathrm{cm}^{-1}$ for the
		optical-only parameters evaluated in the same model. The $g$-constrained
		set is thus the better description of a multiplet it never saw, and one
		that did belong to the least-squares data of Ref.~\cite{wortman1971}. The
		agreement is not automatic: the ground-doublet composition and the
		${}^6H_{13/2}$ splittings depend on the same five parameters through
		different reduced matrix elements.}
	
	\add{Higher in the term the agreement degrades, and it does so
		monotonically: \textcolor{black}{2.6, 6.1 and $21.5~\mathrm{cm}^{-1}$} rms for
		${}^6H_{15/2}$, ${}^6H_{13/2}$ and ${}^6H_{11/2}$. \textcolor{black}{Within
			${}^6H_{11/2}$ the discrepancy is concentrated on a single pair: the model
			leaves the two upper doublets nearly degenerate, at 5961 and
			$5962~\mathrm{cm}^{-1}$, where the fluorescence spectrum resolves them at
			5920 and $5961~\mathrm{cm}^{-1}$. The five remaining comparisons of that
			multiplet agree to better than $6~\mathrm{cm}^{-1}$.} This is the pattern
		expected from the one approximation of the model that is energy-dependent,
		namely the exclusion of the $^6F$ term above $7500~\mathrm{cm}^{-1}$,
		whose second-order influence grows as the multiplet approaches it. The
		optical-only set shows no such trend, which is unsurprising, since all
		these levels were fitted. Above ${}^6H_{11/2}$ no comparison is possible:
		Wortman and Sanders report that term mixing with $^6F$ leaves the
		${}^6H_{9/2}$ and ${}^6H_{7/2}$ positions ambiguous and give no
		experimental assignment for either \cite{wortman1971}. Only two of the
		three ${}^6H_{5/2}$ doublets were located, leaving a single residual once
		the barycenter is removed, so that multiplet carries no statistical
		weight.}
	
	\add{The conclusion relevant to Sec.~\ref{sec:reconciling} is that the
		parameter set is validated where the model is complete and departs from
		experiment only where the model is known to be incomplete. Neither
		behavior bears on the ground doublet, which lies at the opposite end of
		the term from the excluded states.}
	
	\begin{table}
		\caption{\add{Crystal-field levels of the ${}^6H_{13/2}$ and
				${}^6H_{11/2}$ multiplets, observed \cite{wortman1971} and calculated
				in the 66-state model from the optical-only parameters of
				Ref.~\cite{wortman1971} and from the combined fit of
				Eq.~(\ref{eq:Bkq-final}). Neither multiplet entered the combined fit.
				Each calculated set is aligned on the observed barycenter. The two
				lowest ${}^6H_{13/2}$ doublets are unresolved in the fluorescence
				spectrum and both assigned to the line at $3476~\mathrm{cm}^{-1}$.}}
		\label{tab:multiplets}
		\begin{ruledtabular}
			\begin{tabular}{ccc}
				Observed & W\&S \cite{wortman1971} & This work \\
				(cm$^{-1}$) & (optical only) & (Eq.~\ref{eq:Bkq-final}) \\
				\colrule
				\multicolumn{3}{c}{${}^6H_{13/2}$}\\
				3476 & 3486 & \textcolor{black}{3483} \\
				3476 & \textcolor{black}{3492} & \textcolor{black}{3487} \\
				3512 & \textcolor{black}{3512} & \textcolor{black}{3512} \\
				3564 & \textcolor{black}{3557} & \textcolor{black}{3562} \\
				3619 & 3604 & \textcolor{black}{3624} \\
				3697 & 3692 & \textcolor{black}{3689} \\
				3711 & 3708 & \textcolor{black}{3711} \\
				rms  & 9.8  & \textcolor{black}{6.1} \\
				\colrule
				\multicolumn{3}{c}{${}^6H_{11/2}$}\\
				5777 & \textcolor{black}{5767} & \textcolor{black}{5752} \\
				5814 & \textcolor{black}{5815} & \textcolor{black}{5810} \\
				5849 & \textcolor{black}{5863} & \textcolor{black}{5866} \\
				5900 & \textcolor{black}{5893} & \textcolor{black}{5888} \\
				5920 & \textcolor{black}{5940} & \textcolor{black}{5961} \\
				5961 & \textcolor{black}{5948} & \textcolor{black}{5962} \\
				rms  & \textcolor{black}{12.4} & \textcolor{black}{21.5} \\
			\end{tabular}
		\end{ruledtabular}
	\end{table}
	
	\subsection{Results and warnings}
	
	The three lowest doublets of the final solution are
	\begin{equation}
		\begin{aligned}
			E_0 &= 0, \quad g_z = \textcolor{black}{7.22},\ g_\perp = \textcolor{black}{5.45},\\
			\psi_+ &\simeq \textcolor{black}{0.805\,|\tfrac{13}{2}\rangle
				- 0.536\,|{-\tfrac{11}{2}}\rangle
				- 0.205\,|\tfrac{5}{2}\rangle
				- 0.150\,|{-\tfrac{3}{2}}\rangle},\\[4pt]
			E_1 &= \textcolor{black}{3.4}~\mathrm{cm}^{-1}, \quad g_z = \textcolor{black}{5.06},\ g_\perp = \textcolor{black}{6.42},\\
			\psi_+ &\simeq \textcolor{black}{0.834\,|\tfrac{11}{2}\rangle
				+ 0.549\,|{-\tfrac{13}{2}}\rangle},\\[4pt]
			E_2 &= \textcolor{black}{16.3}~\mathrm{cm}^{-1}, \quad g_z = \textcolor{black}{3.2},\ g_\perp = \textcolor{black}{8.4},\\
			\psi_+ &\simeq \textcolor{black}{0.727\,|\tfrac{9}{2}\rangle
				+ 0.503\,|{-\tfrac{7}{2}}\rangle
				+ 0.399\,|\tfrac{1}{2}\rangle
				- 0.245\,|{-\tfrac{15}{2}}\rangle},
		\end{aligned}
		\label{eq:wf}
	\end{equation}
	with weight outside ${}^6H_{15/2}$ below $10^{-3}$ in every case. The
	$M$ content classifies the doublets under $S_4$: the pair at 0 and
	\textcolor{black}{$3.4~\mathrm{cm}^{-1}$} belongs to $\Gamma_{5,6}$, while the doublet at
	\textcolor{black}{$16.3~\mathrm{cm}^{-1}$} is the first level of the opposite
	representation $\Gamma_{7,8}$, the state that governs the
	electric-field couplings of Sec.~\ref{sec:LEFE-origin} through
	Eq.~(\ref{eq:LEFEcross}).
	
	Two warnings apply. The data constrain the parameters along a narrow
	valley rather than at a point: $E_1$ moves between 2 and
	$5~\mathrm{cm}^{-1}$ along this valley, $B_{60}$ is poorly determined,
	and the composition of the ground doublet is the robust output. These
	are effective parameters of the $LS$-coupled $^6H$ model, not values
	transferable to full-configuration calculations.
	
	\section{Estimate of Dy$^{3+}$-Dy$^{3+}$ pair intensity}
	\label{app:pairs}
	
	The low-field satellite lines observed in the $ab$-plane rotation of the
	high-concentration sample (s3) (Fig.~\ref{fig:WF2000ppm}) are attributed
	to exchange- or dipolar-coupled Dy$^{3+}$-Dy$^{3+}$ pairs. We support
	this assignment with a simple combinatorial estimate of the expected
	pair concentration as a function of doping level.
	
	In CaWO$_4$, Dy$^{3+}$ substitutes for Ca$^{2+}$ on the scheelite
	$4b$ site. While the true Ca sublattice is tetragonal, the local
	connectivity of nearest-neighbor cation sites can be reasonably
	approximated, for the purpose of this estimate, by a diamond-type
	cubic lattice, in which each cation site has four nearest cation
	neighbors. Under this approximation, and assuming random, uncorrelated
	substitution of Dy$^{3+}$ for Ca$^{2+}$ with probability $c$ (the
	doping fraction), the probability that a given Dy$^{3+}$ ion
	has at least one Dy$^{3+}$ ion among its four nearest neighbors, and
	therefore forms a pair, is
	\begin{equation}
		P_{\mathrm{pair}}(c) = 1-(1-c)^4 .
		\label{eq:Ppair}
	\end{equation}
	For $c \ll 1$, Eq.~\eqref{eq:Ppair} reduces to $P_{\mathrm{pair}}
	\approx 4c$, i.e., the leading-order probability of finding a pair
	scales linearly with the four available neighbor sites.
	
	To compare with the EPR intensity of a single satellite line, note that
	the four nearest-neighbor directions define four crystallographically
	equivalent pair orientations, which are magnetically inequivalent for a
	general orientation of $H_0$ in the $ab$ plane and therefore resonate
	at distinct fields. Random substitution distributes the paired ions
	equally among the four orientations, so one pair line carries one
	quarter of the total pair intensity. The intensity of a single pair
	resonance relative to that of an isolated, $S_4$-symmetric Dy$^{3+}$
	ion is then
	\begin{equation}
		\frac{I_{\mathrm{pair}}}{I_{\mathrm{single}}} =
		\frac{P_{\mathrm{pair}}(c)}{4} = \frac{1-(1-c)^4}{4}
		\;\xrightarrow{c\ll1}\; c .
		\label{eq:Irel}
	\end{equation}
	Residual factors of order unity are not included in this estimate: the
	near-equal thermal populations of the partner-spin states at 7~K, the
	isotopic composition of the pair, and accidental coincidences of pair
	orientations when $H_0$ lies along a symmetry direction. The estimate
	is therefore reliable at the order-of-magnitude level, which suffices
	for the assignment.
	
	Table~\ref{tab:pairs} gives the values of $P_{\mathrm{pair}}$ and
	$I_{\mathrm{pair}}/I_{\mathrm{single}}$ computed from Eq.~\eqref{eq:Irel}
	for the three samples studied in this work.
	
	\begin{table}[H]
		\centering
		\caption{EPR-calibrated Dy$^{3+}$ concentration, pair formation probability,
			and relative pair-to-single-ion EPR intensity for samples (s1), (s2), and (s3),
			calculated from Eq.~\eqref{eq:Irel}.}
		\label{tab:pairs}
		\begin{tabular}{lccc}
			\hline\hline
			Sample & $c$ & $P_{\mathrm{pair}}$ & $I_{\mathrm{pair}}/I_{\mathrm{single}}$ \\
			\hline
			(s1) & 80~ppb   & $3.2\times10^{-7}$ & $8.0\times10^{-8}$ \\
			(s2) & 2.3~ppm  & $9.2\times10^{-6}$ & $2.3\times10^{-6}$ \\
			(s3) & 247~ppm  & $9.9\times10^{-4}$ & $2.5\times10^{-4}$ \\
			\hline\hline
		\end{tabular}
	\end{table}
	
	The predicted ratio for (s3), $I_{\mathrm{pair}}/I_{\mathrm{single}}
	\approx 2.5\times10^{-4}$, is in quantitative agreement with
	the experimental estimate of a few $10^{-4}$ obtained from the relative
	intensity of the satellite lines with respect to the main $S_4$
	resonance. For (s1) and (s2), the model predicts pair intensities of
	$8\times10^{-8}$ and $2.3\times10^{-6}$, respectively, two to four
	orders of magnitude below the (s3) value and well under the experimental
	noise floor, consistent with the absence of resolved pair satellites in
	the spectra of these two lower-concentration samples (Fig.~\ref{fig:WFPG2}).
	
	This agreement, obtained from a purely combinatorial argument with no
	adjustable parameter, provides strong support for the assignment of the
	low-field satellites in (s3) to Dy$^{3+}$-Dy$^{3+}$ pair centers rather
	than to an alternative single-ion defect mechanism, since the latter
	would not be expected to scale quadratically with doping concentration
	in this manner.
	
	\section{Angular dependence of the Er$^{3+}$ and Dy$^{3+}$ linewidths in the $ab$ plane}
	\label{app:LEFE}
	
	\begin{figure}[h]
		\includegraphics{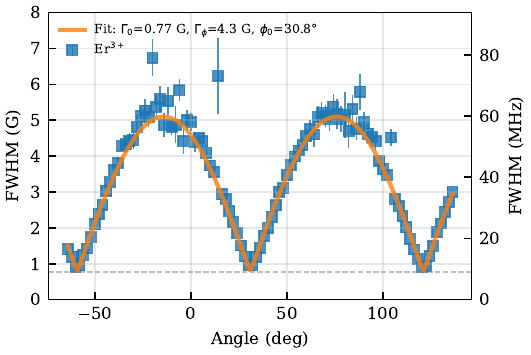}
		\caption{Angular dependence of the Er$^{3+}$ linewidth (FWHM, expressed
			in frequency units using $g_\perp(\mathrm{Er^{3+}}) = 8.38$) in sample
			(s1) for $H_0$ rotated in the $ab$ plane. The solid line is a fit to
			Eq.~(\ref{eq:lefefit}), with $\Gamma_0 = 9.0$~MHz,
			$\Gamma_\varphi = 50$~MHz, and $\varphi_0 = 30.8^\circ$. The dashed
			line indicates $\Gamma_0$.}
		\label{fig:LWErPG2}
	\end{figure}
	
	\begin{figure}[h]
		\includegraphics{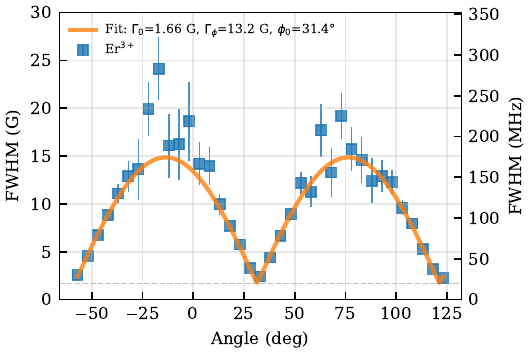}
		\caption{Angular dependence of the Er$^{3+}$ linewidth (FWHM, expressed
			in frequency units using $g_\perp(\mathrm{Er^{3+}}) = 8.38$) in sample
			(s2) for $H_0$ rotated in the $ab$ plane. The solid line is a fit to
			Eq.~(\ref{eq:lefefit}). Best-fit
			parameters: $\Gamma_0 = 19.5$~MHz, $\Gamma_\varphi = 155$~MHz,
			$\varphi_0 = 31.4^\circ$.}
		\label{fig:LWEr10ppm}
	\end{figure}
	
	\begin{figure}[h]
		\includegraphics{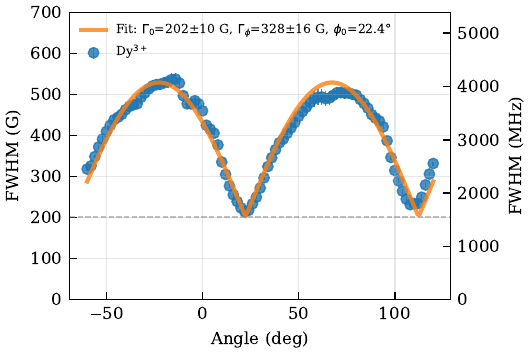}
		\caption{Angular dependence of the Dy$^{3+}$ linewidth (FWHM) for $H_0$
			rotated in the $ab$ plane, in field units (left axis) and in frequency
			units using $g_\perp(\mathrm{Dy^{3+}}) = 5.45$ (right axis) in sample (s3).
			The solid line is a fit to Eq.~(\ref{eq:lefefit}),
			with $\Gamma_0 = 1540$~MHz, $\Gamma_\varphi = 2501$~MHz, and
			$\varphi_0 = 22.4^\circ$. }
		\label{fig:LWDy2000ppm}
	\end{figure}
	
	\section{Group-theoretical origin of the linear and second-order electric field effects}
	\label{app:LEFE-theory}
	
	This appendix derives, from the $S_4$ point-group symmetry of the \ce{Dy^3+} site in \Ca, the selection rules governing the linear electric field effect (LEFE) discussed in Sec.~\ref{sec:LEFE}, following the general perturbative treatment of electric shifts in paramagnetic resonance developed by Kiel~\cite{kiel1966}, and clarifies why the second-order (quadratic) Stark shift remains active under every orientation of $\mathbf{E}$ and $\mathbf{H}_0$ for which the first-order term vanishes.
	
	\subsection{Symmetry of the perturbing field}
	\label{app:LEFE-symmetry}
	
	The electric field Hamiltonian is obtained by expanding the crystal-field potential to first order in an applied (or internal, defect-induced) field $\mathbf{E}$,
	\begin{equation}
		H'_E = \sum_i \frac{\partial H_{\rm CF}}{\partial E_i}\bigg|_{E=0} E_i ,
		\label{eq:HE-expansion}
	\end{equation}
	where $i$ runs over the spherical components $\{+1,-1,z\}$ of $\mathbf{E}$, equivalently $\{E_{+1},E_{-1},E_c\}$ below, matching the $S_4$ irrep decomposition used throughout this appendix, and re-expressing the resulting odd-parity terms as combinations of Stevens operator equivalents $O_k^{(r)}$ acting within the \ce{Dy^3+} ground $J$-multiplet. Because $S_4$ contains no inversion, this expansion is symmetry-allowed to first order; in a centrosymmetric site it would vanish identically.
	
	$S_4$ is Abelian, generated by the improper fourfold rotation about the crystallographic $c$ axis, and has four one-dimensional irreducible representations $A$, $B$, $E$, $E^*$ (the latter two forming a complex-conjugate pair). The field component parallel to $c$, $E_c \equiv E_z$, changes sign under the $S_4$ operation and is therefore odd under $S_4$: it transforms as $B$. The transverse components combine as $E_{\pm1} = \mp(E_x \pm iE_y)/\sqrt{2}$ and transform as the $E,E^*$ pair. No component of $\mathbf{E}$ transforms as the totally symmetric representation $A$; this single fact is the origin of every LEFE selection rule discussed below.
	
	Each Stevens operator $O_k^{(r)}$ shifts the projection quantum number by $\Delta m_J = k$, and its transformation under $S_4$ is fixed by $k \bmod 4$: $A$ for $k \equiv 0$, $B$ for $k \equiv 2$, and $E,E^*$ for $k$ odd. Invariance of $H'_E$ under $S_4$ requires the field component and the paired operator to transform as conjugate representations of the same type. Consequently
	\begin{align}
		H'_E\big|_{E\parallel c} &= E_c \!\!\sum_{r=2,4,6}\!\! \left(R^{(r)}_{02}\,O_2^{(r)} + R^{(r)*}_{02}\,O_{-2}^{(r)}\right) + E_c\left(R^{(6)}_{06}\,O_6^{(6)} + \text{h.c.}\right), \label{eq:HE-par}\\
		H'_E\big|_{E\perp c} &= \!\!\sum_{r=2,4,6}\!\! \left(R^{(r)}_{11}\,E_{+1}\,O_1^{(r)} + \text{h.c.}\right) + \!\!\sum_{r=4,6}\!\! \left(R^{(r)}_{1,-3}\,E_{+1}\,O_{-3}^{(r)} + \text{h.c.}\right) + \left(R^{(6)}_{1,-5}\,E_{+1}\,O_{-5}^{(6)} + \text{h.c.}\right), \label{eq:HE-perp}
	\end{align}
	where the $R^{(r)}_{gk}$ are odd-parity crystal-field coupling coefficients, fixed by the local charge distribution around the \ce{Dy^3+} ion and, in general, ion-dependent: formally, they are defined by re-expanding the field derivative of Eq.~\eqref{eq:HE-expansion} on the Stevens basis, $\partial H_{\rm CF}/\partial E_g \equiv \sum_{r,k} R^{(r)}_{gk}\,O_k^{(r)}$ for each field component $g\in\{+1,-1,z\}$, so that $R^{(r)}_{gk}$ is simply the coefficient of $O_k^{(r)}$ in that expansion. Symmetry restricts the nonzero $R^{(r)}_{gk}$ to the pairings shown in Eqs.~\eqref{eq:HE-par}--\eqref{eq:HE-perp}; their magnitudes are not calculated here and enter only as the ion-dependent constant $K$ of Eq.~\eqref{eq:S1}. A field along $c$ couples exclusively to even-order operators ($k=\pm2,\pm6$); a transverse field couples exclusively to odd-order operators ($k=\pm1,\pm3,\pm5$). Both sets are disjoint from $k=0$: this is the structural statement, valid for any rare-earth ion in this host, any field magnitude, and any orientation of $\mathbf{H}_0$, that no first-order diagonal shift can occur within a pure $\ket{m_J}$ state.
	
	\subsection{First-order effect: role of the spin-Hamiltonian eigenstates}
	\label{app:LEFE-first-order}
	
	The first-order energy correction is a diagonal matrix element,
	\begin{equation}
		\Delta\varepsilon^{(1)}_i = \bra{\psi_i} H'_E \ket{\psi_i} .
		\label{eq:eps1}
	\end{equation}
	Since every operator entering Eqs.~(\ref{eq:HE-par}--\ref{eq:HE-perp}) has $k\neq0$, this vanishes identically whenever $\ket{\psi_i}$ is a pure $\ket{m_J}$ eigenstate, which is precisely the case for $\mathbf{H}_0 \parallel c$, since the Zeeman term $g_\parallel\mu_B H_0 J_z$ commutes with $H_{\rm CF}$ and leaves the $S_4$ crystal-field eigenstates unmixed. No orientation of $\mathbf{E}$ can lift this cancellation. Four configurations follow from combining $\mathbf{E}\parallel c$ or $\mathbf{E}\perp c$ with $\mathbf{H}_0\parallel c$ or $\mathbf{H}_0\perp c$; only the two with $\mathbf{H}_0\perp c$ survive the argument above, and even then not both.
	
	For $\mathbf{H}_0\perp c$, the transverse Zeeman term $g_\perp\mu_B H_0 J_x$ mixes the crystal-field doublet into field-dependent eigenstates $\ket{\psi_i} = \sum_{m_J} c^{(i)}_{m_J}\ket{m_J}$, and diagonal matrix elements of the $k=\pm2$ operators in Eq.~(\ref{eq:HE-par}) become non-zero through cross terms $c^{(i)*}_{m_J} c^{(i)}_{m_J-2}$. This reproduces the Mims result~\cite{mims1964,mims1965} used in Sec.~\ref{sec:LEFE},
	\begin{equation}
		\frac{\partial f}{\partial E_c} = S_1(\varphi)\,f_{\rm EPR}, \qquad S_1(\varphi) = \frac{K}{2g_\perp^2}\sin\!\big[2(\varphi-\varphi_0)\big],
		\label{eq:S1}
	\end{equation}
	with $\varphi$ the azimuth of $\mathbf{H}_0$ in the $ab$ plane, $K$ an ion-dependent constant set by the magnitude of $R^{(r)}_{02}$ and by the energy denominators to the nearest excited crystal-field levels, and $\varphi_0$ the magic angle at which $S_1$ vanishes. The two inversion-related \ce{Ca^2+} sites experience opposite local fields, $E_c^{(2)} = -E_c^{(1)}$, so the two sublattices are shifted by $\pm\delta f^{(1)}$: this is the origin of the field-induced line splitting, symmetric about the unperturbed resonance, that we use in Sec.~\ref{sec:LEFE} to extract $\alpha_{\rm Dy}$ from the residual \ce{Er^3+} probe.
	
	For $\mathbf{E}\perp c$ with $\mathbf{H}_0\perp c$, the $k=\pm1$ operators similarly acquire non-zero diagonal elements once the eigenstates are mixed. Individually, each ion is shifted. But the two inversion-related sites are related not by a simple sign reversal of $E_{\pm1}$ but by a $90^\circ$ rotation of the local $S_4$ frame, and Mims~\cite{mims1964} showed that the resulting shifts cancel exactly on ensemble average. The same vanishing first-order response is therefore obtained for $\mathbf{E}\perp c,\,\mathbf{H}_0\perp c$ as for $\mathbf{E}\perp c,\,\mathbf{H}_0\parallel c$, but by two physically distinct mechanisms: an operator selection rule in one case, and an inversion-site cancellation in the other. This distinction matters for interpreting samples in which the internal field direction is not controlled: a null first-order LEFE signal does not by itself indicate $E\parallel c$ or $E\perp c$.
	
	\subsection{Second-order effect: an always-allowed quadratic shift}
	\label{app:LEFE-second-order}
	
	The second-order correction,
	\begin{equation}
		\Delta\varepsilon^{(2)}_i = \sum_{n\neq i} \frac{|\bra{n} H'_E \ket{\psi_i}|^2}{E_i^{(0)} - E_n^{(0)}},
		\label{eq:eps2}
	\end{equation}
	depends only on the modulus squared of an off-diagonal matrix element. Off-diagonal coupling between the ground doublet and excited crystal-field levels through $H'_E$ is generically non-zero for every orientation of $\mathbf{E}$ considered above: the operators in Eqs.~(\ref{eq:HE-par}--\ref{eq:HE-perp}) that vanish on the diagonal are exactly the ones that connect different $m_J$, hence they act off-diagonally between states separated by $\Delta m_J = k$. There is consequently no orientation of $\mathbf{E}$ or $\mathbf{H}_0$, and no site symmetry lower than centrosymmetric, that suppresses $\Delta\varepsilon^{(2)}$.
	
	Two properties distinguish this term from the first-order shift. First, $(-E)^2 = E^2$, so both inversion-related sites acquire the same second-order correction, of the same sign; the effect does not split the line into a symmetric doublet but instead displaces and skews the entire lineshape asymmetrically, consistent with the Dysonian asymmetry used in Sec.~\ref{sec:epr-params} to fit the linewidth data. Second, because the intermediate states $\ket{n}$ mediating the mixing differ between $\mathbf{H}_0\parallel c$ and $\mathbf{H}_0\perp c$, the sign of the asymmetry parameter can reverse between the two orientations; this reversal is a direct group-theoretical prediction, not a fitted artifact, and provides an independent consistency check on any lineshape asymmetry model applied to the angular-dependent spectra.

	\begin{figure}[h]
		\includegraphics{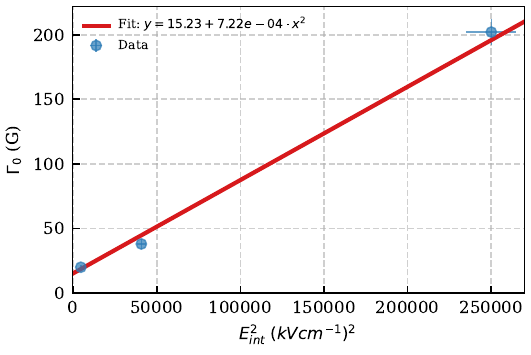}
		\caption{Residual linewidth floor $\Gamma_0$, extracted at
			$\varphi = \varphi_0$ where the first-order LEFE broadening
			vanishes, as a function of the squared internal electric field
			$E_\mathrm{int}^2$ for samples (s1), (s2), and (s3). The solid
			line is a fit to Eq.~\eqref{eq:floor}, yielding
			$\Gamma_{00} = 15.2$~G and
			$\beta = 7.2\times10^{-4}$~G\,(kV/cm)$^{-2}$. The linear
			dependence on $E_\mathrm{int}^2$ is the scaling expected for a
			second-order electric field effect. }
		\label{fig:LEFE2nd}
	\end{figure}
	
	\subsection{Summary}
	
	Table~\ref{tab:LEFEcases} collects the relevant combinations of field and spin orientation. The pattern reduces to one statement: because $S_4$ admits no field component transforming as the totally symmetric representation, the linear Stark shift requires eigenstate mixing to appear at all, and mixing is controlled entirely by the orientation of $\mathbf{H}_0$, not of $\mathbf{E}$. The quadratic shift carries no such restriction and is therefore the only contribution guaranteed to survive in an arbitrary, uncontrolled internal field, the physically relevant situation for charge-compensating defects distributed at random orientations, which is the case treated in the second-order Stark analysis of Sec.~\ref{sec:LEFE2}.
	
	\begin{table}[h]
		\centering
		\begin{tabular}{cc|cc|cc}
			\toprule
			\multicolumn{2}{c|}{Orientation} & \multicolumn{2}{c|}{First order $\Delta\varepsilon^{(1)}$} & \multicolumn{2}{c}{Second order $\Delta\varepsilon^{(2)}$} \\
			$\mathbf{E}$ & $\mathbf{H}_0$ & value & mechanism & value & lineshape \\
			\midrule
			$\parallel c$ & $\parallel c$ & $0$ & selection rule ($k=\pm2,\pm6$) & $\neq0$ & asymmetric \\
			$\parallel c$ & $\perp c,\ \varphi\neq\varphi_0$ & $\neq0$ & mixing + $k=\pm2$ & $\neq0$ & split + asymmetric \\
			$\parallel c$ & $\perp c,\ \varphi=\varphi_0$ & $0$ & $\sin[2(\varphi-\varphi_0)]=0$ & $\neq0$ & asymmetric \\
			$\perp c$ & $\parallel c$ & $0$ & selection rule ($k=\pm1,\pm3,\pm5$) & $\neq0$ & asymmetric \\
			$\perp c$ & $\perp c$ & $0$ & inversion-site cancellation & $\neq0$ & asymmetric \\
			\bottomrule
		\end{tabular}
		\caption{First- and second-order LEFE for the relevant orientations of $\mathbf{E}$ and $\mathbf{H}_0$ in $S_4$ symmetry. The second-order effect is active in every case, with a sign that can differ between $\mathbf{H}_0\parallel c$ and $\mathbf{H}_0\perp c$.}
		\label{tab:LEFEcases}
	\end{table}
	
\end{document}